\documentclass[twocolumn,showpacs,preprintnumbers,amsmath,amssymb]{revtex4}
\usepackage{graphicx}
\usepackage{dcolumn}
\usepackage{bm}%

\newtheorem{thm}{Theorem}
\newtheorem{cor}{Corollary}
\newtheorem{defi}{Definition}

\newtheorem{lem}{Lemma}
\newtheorem{rem}{Remark}

\newenvironment{proof}[1][Proof]{\textbf{#1.} }{\ \rule{0.5em}{0.5em}}

\newcommand{\C}{\mathbb C}
\newcommand{\N}{\mathbb N}

\newcommand{\R}{\mathbb R}

\def\la{\label}
\def\bt{\begin{thm}}
\def\et{\end{thm}}

\def\bl{\begin{lem}}
\def\el{\end{lem}}

\def\bd{\begin{defi}}
\def\ed{\end{defi}}

\def\bc{\begin{cor}}
\def\ec{\end{cor}}

\def\bp{\begin{proof}}
\def\ep{\end{proof}}

\def\br{\begin{rem}}
\def\er{\end{rem}}

\begin{document}

\title{Superfluidity of Helium-3}

\thanks{The work was supported in part by the
Office of Naval Research and by the National Science Foundation.}
\author{Tian Ma}
\affiliation{Department of Mathematics, Sichuan University,
Chengdu, P. R. China}%

\author{Shouhong Wang}
 \homepage{http://www.indiana.edu/~fluid}
\affiliation{Department of Mathematics,
Indiana University, Bloomington, IN 47405}%
\date{\today}

\begin{abstract}
This article presents a phenomenological dynamic phase transition theory -- modeling and analysis -- for liquid helium-3. We derived two new models, for liquid helium-3 with or without applied field,  by introducing three wave functions and using a unified dynamical Ginzburg-Landau model. The analysis of these new models leads to predictions of existence of 1) a unstable region, 2) a new phase C in a narrow region, and 3) switch points of transition types on the coexistence curves near two triple points.
It is hoped that these predictions 
  will be useful for designing better physical experiments and lead to better understanding of the physical mechanism of superfluidity.
\end{abstract}

\maketitle
\section{Introduction}
\label{sc1}
Superfluidity is a phase of matter  in which "unusual" effects are observed when liquids, typically of helium-4 or helium-3, overcome friction by surface interaction when at a stage, known as the "lambda point" for helium-4, at which the liquid's viscosity becomes zero. Experiments have indicated that helium atoms have two stable isotopes $^4$He and $^3$He.  $^3$He contains two electrons, two
protons and one neutron. Hence it has a fractional spin and obey
the Fermi-Dirac statistics. The liquid $^3$He has two types of superfluid
phases: phase A and  phase  B. In particular, if we apply a
magnetic field on the liquid $^3$He, then there will be a 
third superfluid phase, called the $A_1$ phase. 

The main objectives  of this article are 1) to establish some dynamical 
Ginzburg-Landau models  for $^3$He with or without applied magnetic fields, and 2) to study superfluid dynamic transitions and their physical significance.  

Consider the case without applied magnetic field. 
The  modeling is based on two  main ingredients as follows. 

First,  instead of using a single wave function $\psi$, we use three wave functions  to represent Anderson-Brinkman-Morel (ABM) state  and the Balian-Werthamer (BM)  state.  More precisely, we introduce three complex valued
functions $\psi_0, \psi_1, \psi_2$ to characterize the
superfluidity of $^3$He , with  $\psi_0$ for the state
$|\uparrow\uparrow>$,  $\psi_1$  for the state
$|\downarrow\downarrow>$, and $\psi_2$  for the state
$|\uparrow\downarrow >+|\downarrow\uparrow>$. 
Then we are able  to formulate a Ginzburg-Landau (GL) energy in terms of these three wave functions and the density function  $\rho_n$ of the normal fluid state.

Second, we use  a unified  time-dependent Ginzburg-Landau model for equilibrium phase transitions, developed recently by the authors \cite{MW08c, MW08f}, to derive a general time-dependent GL model for $^3$He.

In a nutshell,  the model is obtained by a careful examination of  the classical phase transition diagrams and by using  both mathematical and physical insights offered by a recently  developed  dynamical transition theory as briefly recalled in the appendix.
The model of the case with applied magnetic field can be derived 
in the same fashion.

\medskip

With the models in our disposal, we can study the dynamic phase  transitions of liquid $^3$He, and derive some physical  predictions.

To be precise,  we first recall the classical phase transition diagrams of $^3$He, as shown in Figure~\ref{f8.38}; see among many others  
Ginzburg \cite{ginzburg}, Reichl \cite{reichl}  and  Onuki \cite{onuki}. 
\begin{figure}
  \centering
  \includegraphics[width=0.2\textwidth]{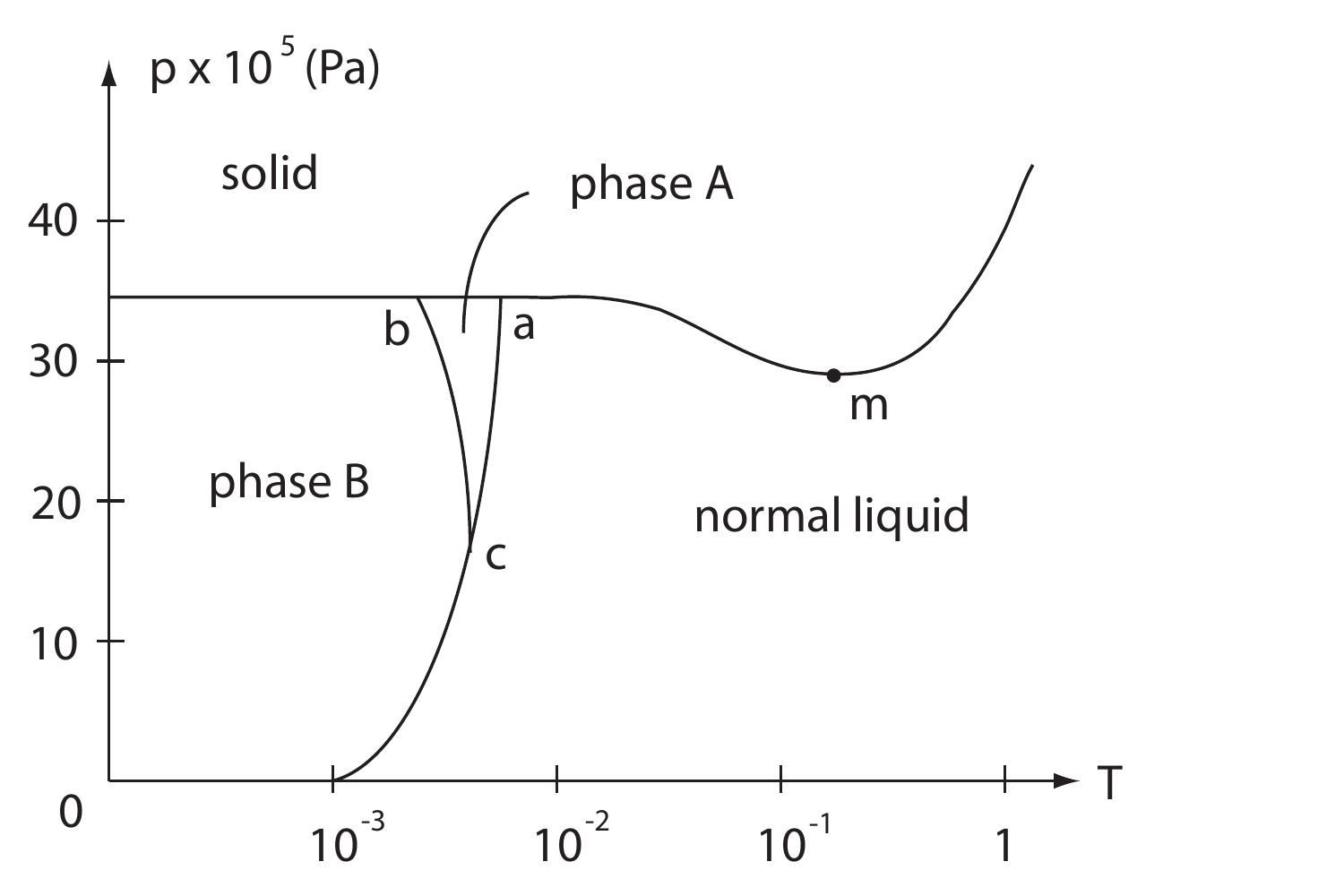}
  \caption{The coexistence curve of $^3$He  in the case
without an applied magnetic field.}\la{f8.38}
 \end{figure}
From this diagram, we see that there  are two coexistence curves, with one separating the solid state and the phase $A$ and phase $B$  superfluid states, and the other separating the superfluid states and the normal liquid state. The transition crossing the 
coexistence curve between the solid and superfluid states is first order (Type-II in the sense of dynamic classification scheme given in the appendix), and the transition crossing the coexistence curve between the superfluid and normal liquid states are second order (Type-I in the sense of dynamic classification scheme).  In addition, the  second transition between phase A and phase B superfluid states is first order (Type-II).

We would like to mention that the Ginburg-Landau theory with only one wave function can not describe this phase transition diagram, and this is one of the mains reasons that we need a new Ginzburg-Landau model as discussed above  to  study  the phase transition dynamics for $^3$He.

The models established are analyzed using a recently developed dynamical transition theory, leading to some interesting physical predictions. 
Here we address briefly the new results derived. For the case without the applied magnetic field, the main results obtained are synthesized in  a theoretical  $PT$-phase diagram of $^3$He given by  Figure~\ref{f8.41}.
\begin{figure}[hbt]
  \centering
  \includegraphics[width=0.3\textwidth]{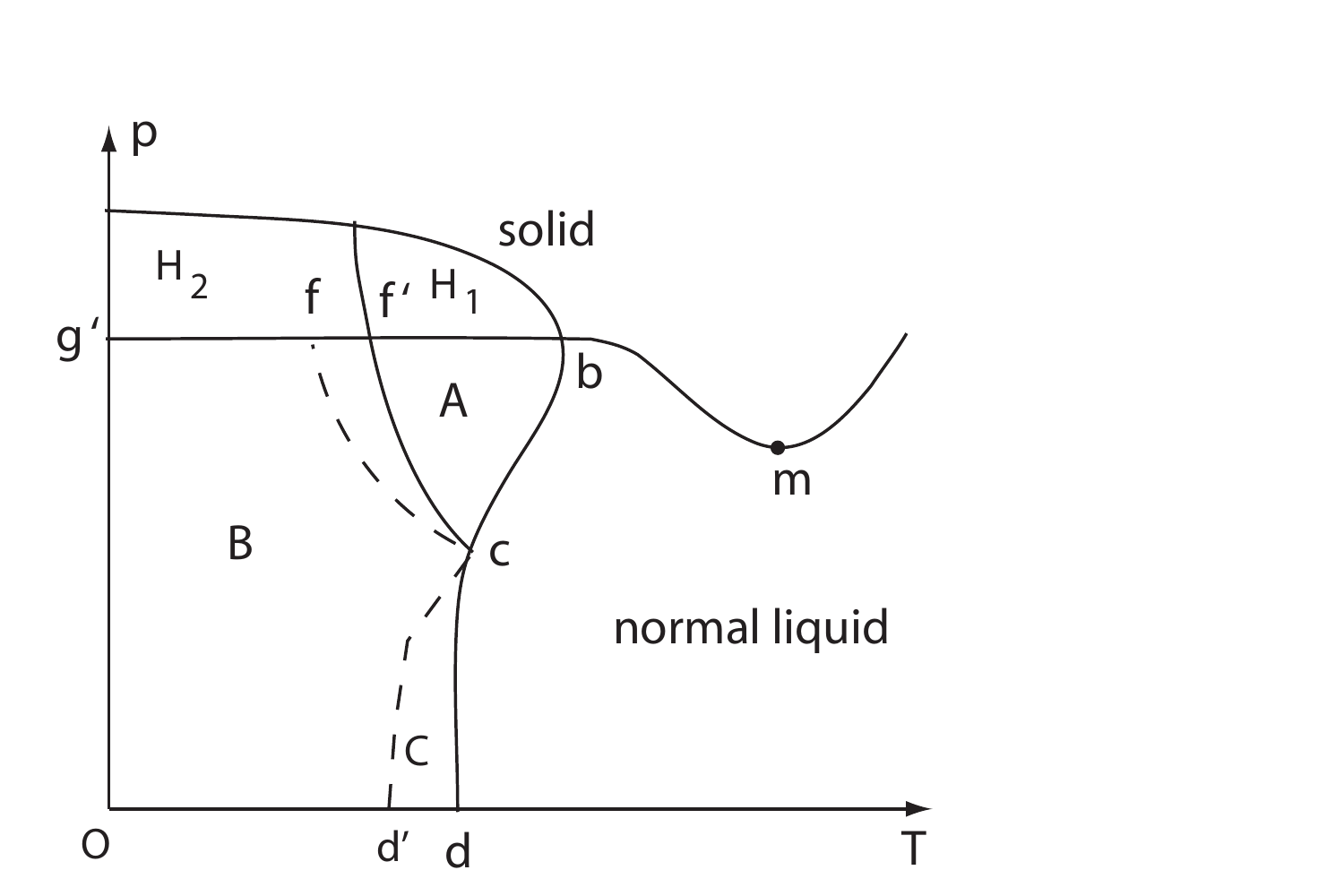}
  \caption{Derived theoretical  $PT$-phase diagram of $^3$He: 
The region $H=H_1 \cup H_2$ is the unstable domain, where the solid state and the superfluid state appear randomly depending on fluctuations. The curve $\widehat{bcd}$ is the first critical curve where phase transition between normal fluid and superfluid states occur.}
\la{f8.41}
 \end{figure}
 
One prediction from our results is  the existence of a unstable region  $H=H_1 \cup H_2$, in which  the solid state and the superfluid states $A$ and $B$ appear randomly depending on fluctuations.   In particular, in $H_1$, phase $B$ superfluid state and the solid may appear, and in $H_2$ the phase $A$ superfluid state and the solid sate may appear. 

Another prediction is the possible  existence of phase C superfluid state, which is characterized by the wave function $\psi_2$, representing $|\uparrow\downarrow >+|\downarrow\uparrow>$. However, phase C region is very narrow, which may be the reason why it is hard to be observed in experiments. 
 
Also, the results predict  that near the two triple points $b$  and $c$, there is a possibility of the existence of two switch points, where the transition on the corresponding coexistence curve switches types at each switch point. The existence of such switch points depends on the physical parameters.

One important new ingredient for the analysis is  a dynamic transition theory developed recently by the authors \cite{chinese-book, b-book}. With this theory, we derive a  new dynamic phase transition classification scheme, which   classifies phase transitions into three categories:  Type-I, Type-II and Type-III, corresponding respectively to the continuous, the jump and mixed transitions in the dynamic transition theory.

The case with applied field can be addressed in the similar fashion; see Section V.

This article is organized as follows. Section II introduces a  new dynamic model for liquid helium-m without applied magnetic field. The phase transition dynamics is given in Section III, and the type of transitions in different regimes are given in Section IV. 
Section V deals with the case with applied magnetic field, and Section VI gives a summary on the physical predications of the new models.

 \section{Dynamic Model for Liquid $^3$He with Zero Applied Magnetic Field}
 The superfluidity of liquid $^3$He was found in 1971 by D. M. Lee,
 D. D. Osheroff, and R. C. Richardson, and its transition temperature
 is $T\simeq 10^{-3}K$ under $p=1$ atm $(10^5$Pa). There are two
 superfluid phases $A$ and $B$ if there is no applied magnetic field.
 Figure \ref{f8.38} provides the phase diagram of the liquid $^3$He in the 
 $PT$-plane.

Because the atoms $^3$He are fermions, to form the superfluid
phase they must be paired to become bosons. When no magnetic field
is applied, there are two pairing states: the
Anderson-Brinkman-Morel  (ABM) state and the Balian-Werthamer (BW) state, given respectively by 
\begin{align*}
& \sqrt{2}|\Phi >=(a+ib)|\ \uparrow\uparrow
>+(a-ib)|\downarrow\downarrow>,
\\
&
\sqrt{2}|\Phi
> = (a+ib)|\uparrow\uparrow>+(a-ib)|\downarrow\downarrow>
+c[|\uparrow\downarrow>+|\downarrow\uparrow>].
\end{align*}
The $ABM$ state corresponds to the superfluid phase $A$, and the
$BW$ state to the phase $B$. Also, we call state 
$|\uparrow\downarrow>+|\downarrow\uparrow>$  as phase $C$, which appears in the 
theory developed in this article.

We introduce three complex valued
functions $\psi_0, \psi_1, \psi_2$ to characterize the
superfluidity of $^3$He , in which $\psi_0$ to the state
$|\uparrow\uparrow>, \psi_1$ to the state
$|\downarrow\downarrow>$, and $\psi_2$ to the state
$|\uparrow\downarrow >+|\downarrow\uparrow>$.

Let $\rho_n$ be the normal fluid density,  $\rho_a$, 
$\rho_b$  and $\rho_c$  represent the densities of superfluid phases $A$, $B$  and $C$ respectively.
Then we have
\begin{align*}
&\rho_a=\tau_0|\psi_0|^2+\tau_1|\psi_1|^2 && (\tau_0>0,\tau_1>0),\\
&\rho_b=\tau_2|\psi_0|^2+\tau_3|\psi_1|^2+\tau_4|\psi_2|^2  && 
(\tau_2,\tau_3 >  0,\tau_4>0), \\
& \rho_c = \tau_5 |\psi_2|^2 && (\tau_5 >0).
\end{align*}
The total density of $^3$He is given by
$$\rho =\left\{\begin{aligned}
& \rho_n             &&\text{ in the  normal state},\\ 
& \rho_n+\rho_a &&\text{ in the phase A state}, \\
& \rho_n+\rho_b &&\text{ in the phase B state},\\
& \rho_n+\rho_c &&\text{ in the phase C state}.
\end{aligned}
\right.$$

Physically, the states $\psi_0,\psi_1$ and $\psi_2$ are
independent, and consequently  there are no  coupling  terms $|\nabla
(\psi_i+\psi_j)|^2$ and $|\psi_i+\psi_j|^{2k}$  $(i\neq j)$ in  the free
energy density. Since in the case without applied magnetic field
$\psi_0$ and $\psi_1$ are equal, their coefficients in the free energy
should be the same. Thus, we have  the  following 
Ginzburg-Landau free energy for $^3$He with $H=0$:
\begin{widetext}
\begin{align}
G(\psi_0,\psi_1,\psi_2,\rho_n)
=& \frac{1}{2}\int_{\Omega}\Big[\frac{k_1h^2}{m}|\nabla\psi_0|^2+
\alpha_1|\psi_0|^2+\alpha_2\rho_n|\psi_0|^2  
+\frac{\alpha_3}{2}|\psi_0|^4 +\frac{k_1h^2}{m}|\nabla\psi_1|^2+\alpha_1|\psi_1|^2 
\nonumber \\
& +\alpha_2 \rho_n|\psi_1|^2+\frac{\alpha_3}{2}|\psi_1|^4 
  +\frac{k_2h^2}{m}|\nabla\psi_2|^2+\beta_1|\psi_2|^2+\frac{\beta_2}{4}|\psi_2|^4 +\beta_3|\psi_0|^2|
\psi_2|^2\nonumber \\
&+\beta_3|\psi_1|^2|\psi_2|^2
+\beta_4\rho_n|\psi_2|^2  +k_3|\nabla\rho_n|^2+\mu_1|\rho_n|^2+\frac{2\mu_2}{3}\rho^3_n+\frac{\mu_3}{2}
\rho^3_n -p\left(\rho_n+\frac{\mu_0}{2}\rho^2_n\right)\Big]dx, \label{8.223}
\end{align}
\end{widetext}
where the coefficients depends on $T$ and $p$, and for $1\leq i\leq 3$,  $j=2,3,4$, 
\begin{equation}
\left.
\begin{aligned} 
&  k_i>0,  \qquad  \beta_j>0, \\
& \alpha_2> 0,\ \ \ \ \alpha_3>0,\ \ \ \ \mu_3>0,\ \ \ \ \mu_2<0.
\end{aligned}
\right.\label{8.224}
\end{equation}
For $\alpha_1,\beta_1$ and $\mu_1$, there are regions $A_i, B_i,
C_i$  $(i=1,2)$ in   the $PT$-plane $\R^2_+$ such that
$\bar{A}_1+\bar{A}_2=\bar{B}_1+\bar{B}_2=\bar{C}_1+\bar{C}_2=\R^2_+$, and
\begin{align}
& 
\alpha_1=\alpha_1(T,p)\left\{\begin{array}{ll}
>0 & \text{ if } (T,p)\in A_1,\\
<0 & \text{ if }(T,p)\in A_2,
\end{array}
\right.\label{8.225}
\\
&
\beta_1=\beta_1(T,p)\left\{\begin{array}{ll}
>0 & \text{ if } (T,p)\in B_1,\\
<0&  \text{ if } (T,p)\in B_2,
\end{array}
\right.\label{8.226}
\\
&
\mu_1=\mu_1(T,p)\left\{\begin{array}{ll}
>0 & \text{ if }  (T,p)\in C_1,\\
<0 & \text{ if }(T,p)\in C_2.
\end{array}
\right.\label{8.227}
\end{align}

It is known that for $^3$He , $\mu_1=\mu_1(T,p)$ is not monotone on
$T$. In fact, at $T_m=0.318K, p_m=29.31\times 10^5$Pa, we have
\begin{equation}
\mu_1(T_m,p_m)=0,\ \ \ \ \frac{\partial\mu_1(T_m,p_m)}{\partial
T}=0,\label{8.228}
\end{equation}
where $m=(T_m,p_m)$ is as shown in Figure \ref{f8.38}. Near  the  point $m$ the
famous Pomeranchuk effect takes place, i.e., when pressure
increases, the liquid $^3$He will absorb heat to undergo a transition to
solid state.

By the normalized model (\ref{7.30}), we infer from (\ref{8.223}) the following time-dependent GL  equations for  the superfluidity of liquid $^3$He :
\begin{widetext}
\begin{equation}
\begin{aligned} 
&\frac{\partial\psi_0}{\partial
t}=\frac{k_1h^2}{m}\Delta\psi_0-\alpha_1\psi_0-\alpha_2\rho_n\psi_0-\beta_3|\psi_2|^2\psi_0-\alpha_3
|\psi_0|^2\psi_0,\\
&\frac{\partial\psi_1}{\partial
t}=\frac{k_1h^2}{m}\Delta\psi_1-\alpha_1\psi_1-\alpha_2\rho_n\psi_1-\beta_3|\psi_2|^2\psi_1-\alpha_3|\psi_1|^2\psi_1,\\
&\frac{\partial\psi_2}{\partial
t}=\frac{k_2h^2}{m}\Delta\psi_2-\beta_1\psi_2-\beta_3|\psi_0|^2\psi_2-\beta_3|\psi_1|^2\psi_2-\beta_4
\rho_n\psi_2-\beta_2|\psi_2|^2\psi_2,\\
&\frac{\partial\rho_n}{\partial
t}=k_3\Delta\rho_n-(\mu_1-\mu_0p)\rho_n-\mu_2\rho^2_n-\mu_3\rho^3_n-\frac{\alpha_2}{2}|\psi_0|^2-
\frac{\alpha_2}{2}|\psi_1|^2   -\frac{\beta_4}{2}|\psi_2|^2-p.
\end{aligned}
\label{8.229}
\end{equation}

The nondimensional form of (\ref{8.229}) can be written as
\begin{equation}
\begin{aligned} 
&\frac{\partial\psi_0}{\partial
t}=\Delta\psi_0+\lambda_1\psi_0-a_1\rho_n\psi_0-a_2|\psi_2|^2\psi_0-a_3|\psi_0|^2\psi_0,\\
&\frac{\partial\psi_1}{\partial
t}=\Delta\psi_1+\lambda_1\psi_1-a_1\rho_n\psi_1-a_2|\psi_2|^2\psi_1-a_3|\psi_1|^2\psi_1,\\
&\frac{\partial\psi_2}{\partial
t}=\kappa_1\Delta\psi_2+\lambda_2\psi_2-b_1\rho_n\psi_2-b_2|\psi_0|^2\psi_2-b_2|\psi_1|^2\psi_2-b_3|
\psi_2|^2\psi_2,\\
&\frac{\partial\rho_n}{\partial
t}=\kappa_2\Delta\rho_n+\lambda_3\rho_n-c_1|\psi_0|^2-c_1|\psi_1|^2-c_2|\psi_2|^2+c_3\rho^2_n-c_4 \rho^3_n,
\end{aligned}
\label{8.230}
\end{equation}
\end{widetext}
where
\begin{equation}
\label{8.231}
\begin{aligned}
&\lambda_1=-\frac{ml^2}{h^2k_1}(\alpha_1+\alpha_2\rho^0_n),\\
&\lambda_2=-\frac{ml^2}{h^2k_1}(\beta_1+b_1\rho^0_n),\\
&\lambda_3=-\frac{ml^2}{h^2k_1}(\mu_1-\mu_0p-2\rho^0_n\mu_2-3(\rho^0_n)^2\mu_3),
\end{aligned}
\end{equation}
and  $\rho^0_n$ is determined by the state state solution of the form
$(\psi_0, \psi_1, \psi_2, \rho_n)=(0, 0, 0, \rho_n^0)$ of (\ref{8.229}).
By (\ref{8.224}) the
coefficients in (\ref{8.230}) satisfy
$$a_i>0,\ \ \ \ b_i>0,\ \ \ \ c_j>0\qquad \forall 1\leq i\leq 3,\ \ \ \
1\leq j\leq 4.$$

The boundary conditions are the Neumann conditions
\begin{equation}
\frac{\partial }{\partial n}  (\psi_0, \psi_1, \psi_2, \rho_n) =0\qquad \text{ on } \partial \Omega. \label{8.232}
\end{equation}

When $p$ is a constant, the problem (\ref{8.230}) with
(\ref{8.232}) can be reduced to the following system of  ordinary differential
equations:
\begin{equation}
\left.
\begin{aligned}
&\frac{d\rho_1}{dt}=\lambda_1\rho_1-a_1\rho_n\rho_1-a_2\rho_2\rho_1-a_3\rho^2_1,\\
&\frac{d\rho_2}{dt}=\lambda_2\rho_2-b_1\rho_n\rho_2-b_2\rho_1\rho_2-b_3\rho^2_2,\\
&\frac{d\rho_n}{dt}=\lambda_3\rho_n-c_1\rho_1-c_2\rho_2+c_3\rho^2_n-c_4\rho^3_n,
\end{aligned}
\right.\label{8.233}
\end{equation}
where $\rho_1=|\psi_0|^2+|\psi_1|^2 $ and $ \rho_2=|\psi_2|^2.$

\section{Critical parameter curves and $PT$-phase diagram}

\subsection{Critical parameter curves}
Critical parameter curves in the $PT$-plane are given by
$$l_i=\{(T,p)\in \R^2_+|\ \lambda_i(T,p)=0\},\ \ \ \ i=1,2,3,$$
where $\lambda_i=\lambda_i(T,p)$ are defined by  (\ref{8.231}).

It is clear that the critical parameter curves $l_i$ are
associated with the $PT$-phase diagram of $^3$He . As in  the last 
section 
if we can determine the critical parameter
curves $l_i$  $(1\leq i\leq 3)$, then we obtain the $PT$-phase diagram.

Phenomenologically, according to the experimental $PT$-phase
diagram (Figure \ref{f8.38}), the parameter curves $l_i$ $(i=1,2,3)$ in the 
$PT$-plane should be  as schematically illustrated in Figure
\ref{f8.39}(a)-(c). The combination of the diagrams (a)-(c) in Figure \ref{f8.39}
gives Figure \ref{f8.40}, in which the real line $\widehat{bm}$ stands
for the coexistence curve of the solid and liquid phases, and
$\widehat{bcd}$ for the coexistence curve of superfluid and normal
liquid phases.
\begin{figure}[hbt]
  \centering
  \includegraphics[width=0.17\textwidth]{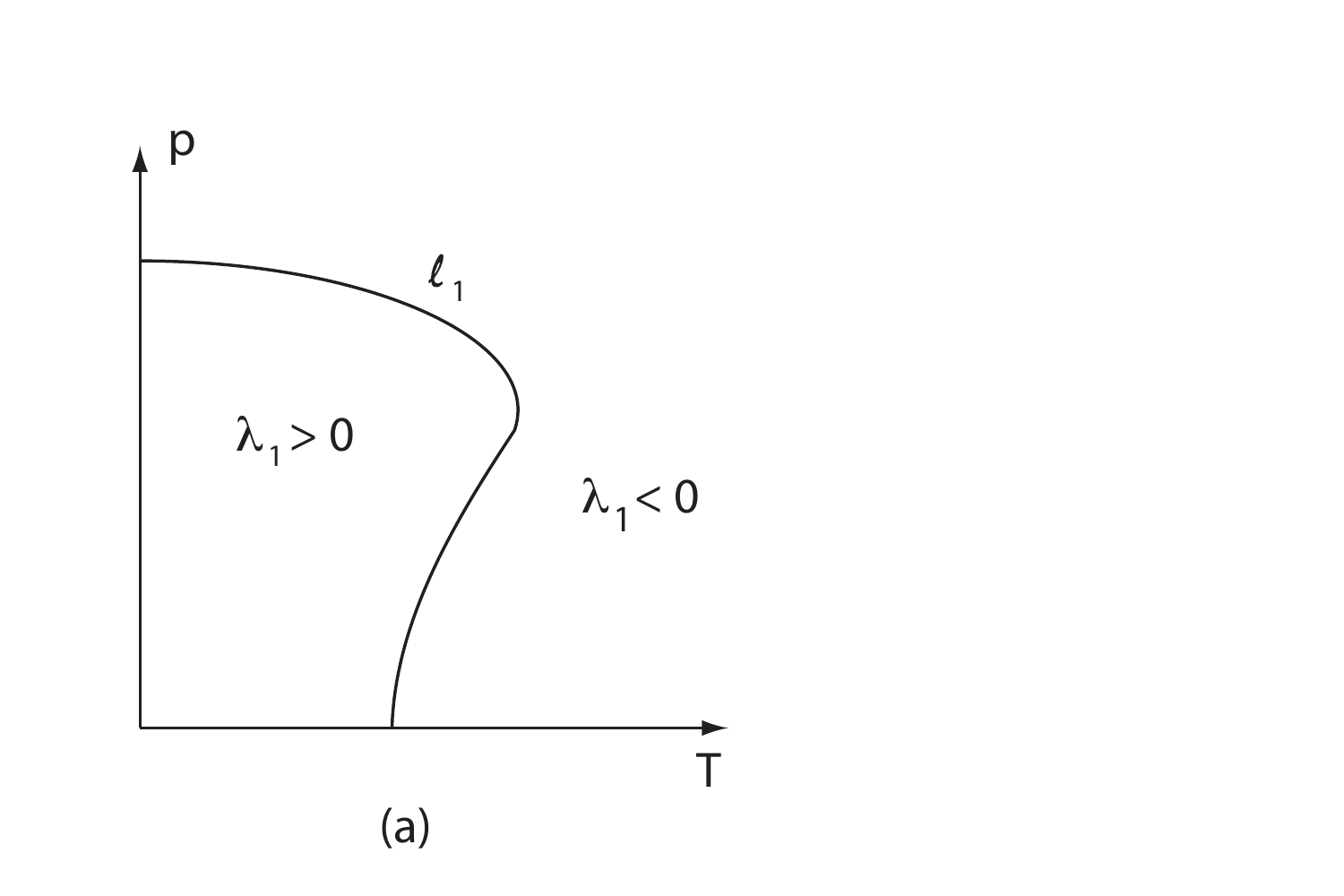} 
  \includegraphics[width=0.17\textwidth]{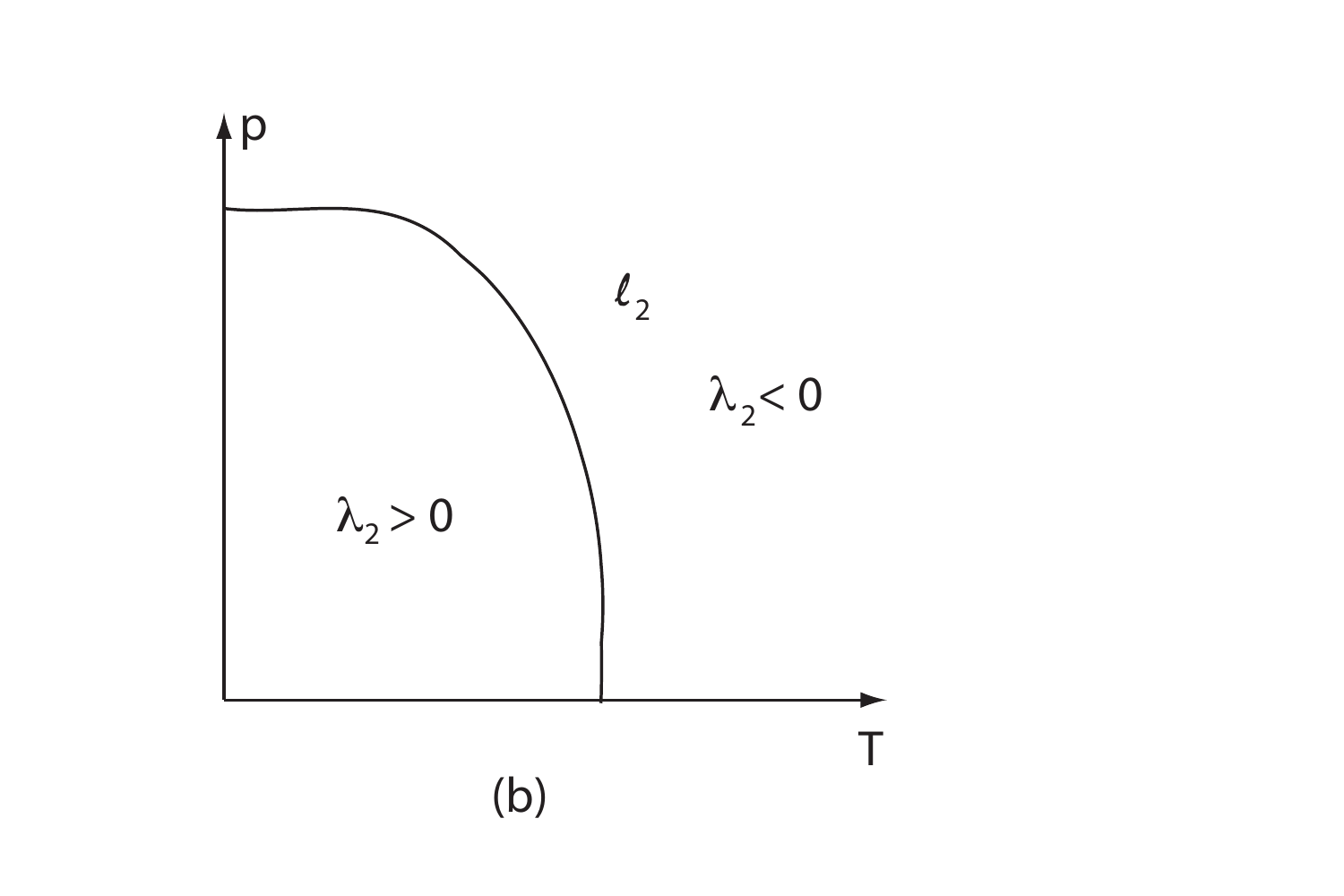}
  \includegraphics[width=0.2\textwidth]{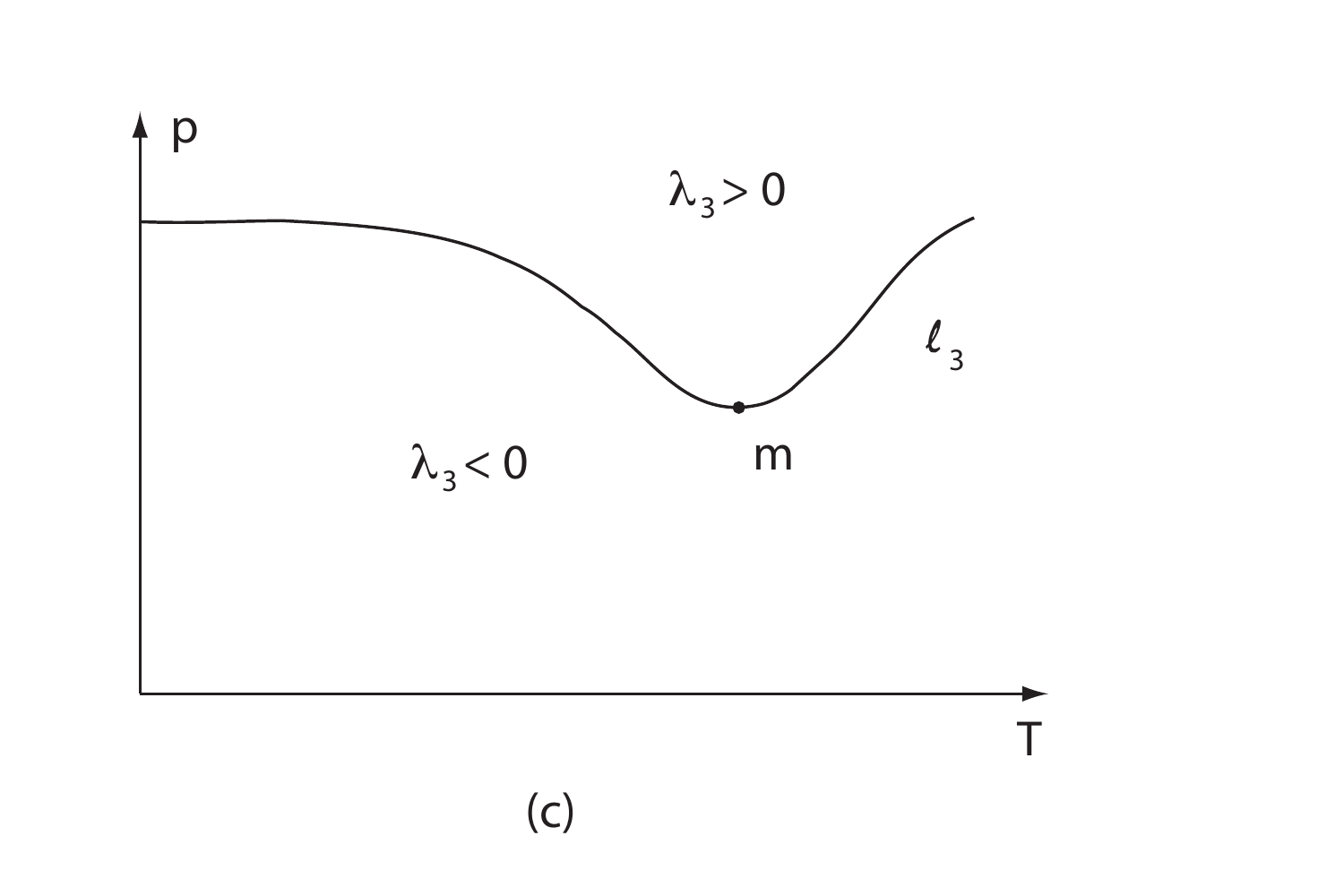}
  \caption{(a) The curve $l_1$  $(\lambda_1=0)$, (b) the
curve $l_2$ $(\lambda_2=0)$,   and (c) the curve $l_3 $ $(\lambda_3=0)$.}\la{f8.39}
 \end{figure}
 
 \begin{figure}[hbt]
  \centering
  \includegraphics[width=0.2\textwidth]{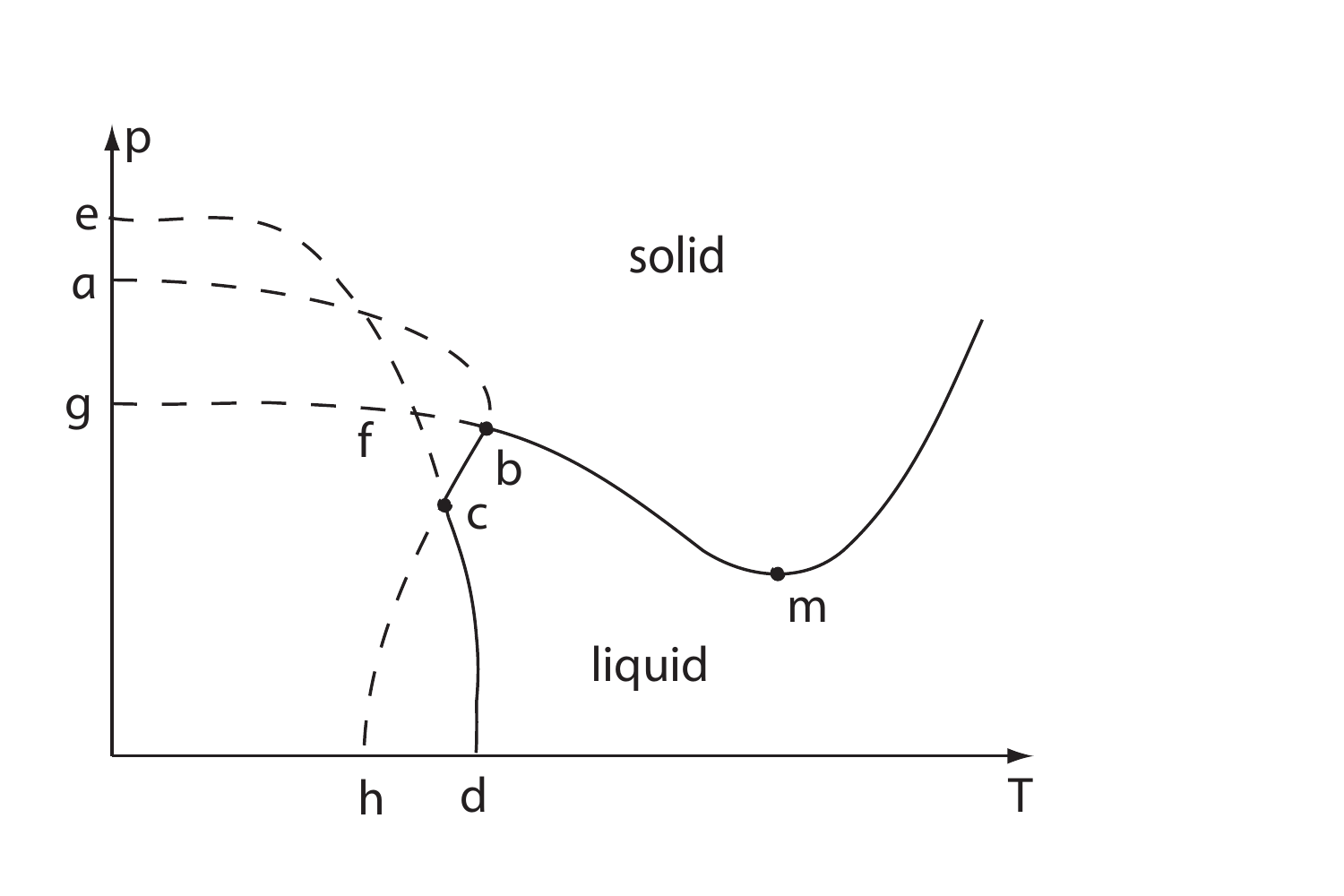}
  \caption{The curve $\widehat{abch}$ is $l_1$,
$\widehat{efcd}$ is $l_2$, and $\widehat{gfbm}$ is $l_3$.}\la{f8.40}
 \end{figure}
 
 Now we rigorously examine phase transitions in different regimes determined by the equations and the critical parameter curves.

\subsection{States in the unstable region}
 
 We consider the dynamical properties of transitions for
(\ref{8.233}) in the unstable region. It is clear that at point
$b=(T_b,p_b)$,
\begin{equation}
\lambda_1(T_b,p_b)=\lambda_3(T_b, p_b)=0,\ \ \ \ \lambda_2(T_b,p_b)<0,\label{8.219}
\end{equation}
and the unstable region $H_1$ near the triple point $b$ is defined by 
$$H_1=\{(T,p)\in \R^2_+\ |\  (\lambda_1,\lambda_2, \lambda_3)(T,p)=(+, -, +)\}.$$

To study the structure of flows of (\ref{8.233}) for $(T,p)\in H$
it is necessary to consider the equations (\ref{8.233}) at the  point
$b=(T_b,p_b)$, and by (\ref{8.219}) which are given by
\begin{equation}
\left.
\begin{aligned}
&\frac{d\rho_1}{dt}=-a_1\rho_n\rho_1-a_3\rho^2_1,\\
&\frac{d\rho_n}{dt}=-c_1\rho_1+c_3\rho^2_n-c_4\rho^3_n,
\end{aligned}
\right.\label{8.233-1}
\end{equation}

We know that  
\begin{equation}
c_3 > 0\ \ \ \ \text{for}\ \ \ \ (T,p)\subset H_1.\label{8.221}
\end{equation}
Equations (\ref{8.233-1}) have
the following two steady state solutions:
\begin{eqnarray*}
&&Z_1=(\rho_1,\rho_n)=(0, c_3/a_4),\\
&&Z_2=(\rho_1,\rho_n)=\left(\frac{a_1}{a_3}\alpha ,-\alpha
\right),\\
&&\alpha =\frac{c_3}{2c_4}\left(\sqrt{1+\frac{4c_1a_1c_4}{a_3 c_3^2}}-1\right).
\end{eqnarray*}
By direct computation, we can prove that the eigenvalues of the
Jacobian matrices of (\ref{8.233-1}) at $Z_1$ and $Z_2$ are
negative. Hence, $Z_1$ and $Z_2$ are stable equilibrium points of
(\ref{8.233-1}). Physically, $Z_1$ stands for solid state, and $Z_2$
for superfluid state. The topological structure of (\ref{8.233-1})
is schematically illustrated by Figure \ref{f8-37-1}(a), the two regions
$R_1$ and $R_2$ divided by curve $AO$ in Figure \ref{f8-37-1}(b) are the
basins of attraction of $Z_1$ and $Z_2$ respectively.

We note that in $H_1$, $ \lambda_1$ and $\lambda_3$ are  small, i.e.,
$$0<\lambda_1(T,p),\ \ \ \ \lambda_2(T,p)\ll 1,\ \ \ \ \text{for}\
(T,p)\in H,
$$ 
and (\ref{8.233-1}) can be considered as a perturbed
system of (\ref{8.233}).

Thus, for $(T,p)\in H_1$ the system (\ref{8.233}) have four steady
state solutions $\widetilde{Z}_i=\widetilde{Z}(T,p)$  $(1\leq i\leq
4)$ such that
$$\lim_{(T,p)\to (T_C,p_C)}   ( \widetilde{Z}_1,   \widetilde{Z}_2, 
\widetilde{Z}_3, \widetilde{Z}_4)(T,p)=(
 Z_1,  Z_2, 0,  0),$$ 
 and $\widetilde{Z}_1$ and
$\widetilde{Z}_2$ are stable, representing solid state and liquid
He-3 state respectively, $\widetilde{Z}_3$ and $\widetilde{Z}_4$
are two saddle points. The topological structure of (\ref{8.233-1})
for $(T,p)\in H_1$ is schematically shown in Figure \ref{f8-37-1}(c), and the
basins of attraction of $\widetilde{Z}_1$ and $\widetilde{Z}_2$
are $\widetilde{R}_1$ and $\widetilde{R}_2$ as illustrated by
Figure \ref{f8-37-1}(d).
\begin{figure}[hbt]
  \centering
  \includegraphics[width=0.18\textwidth]{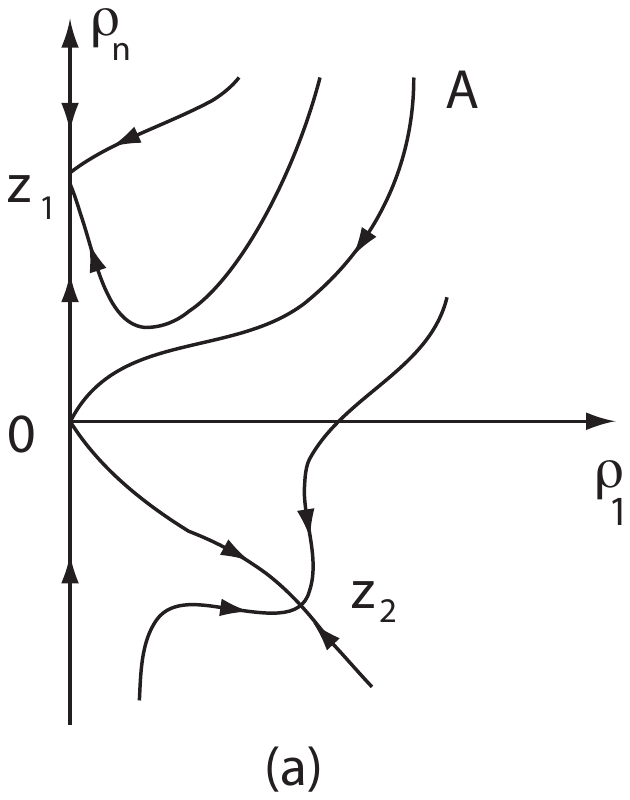}
  \includegraphics[width=0.2\textwidth]{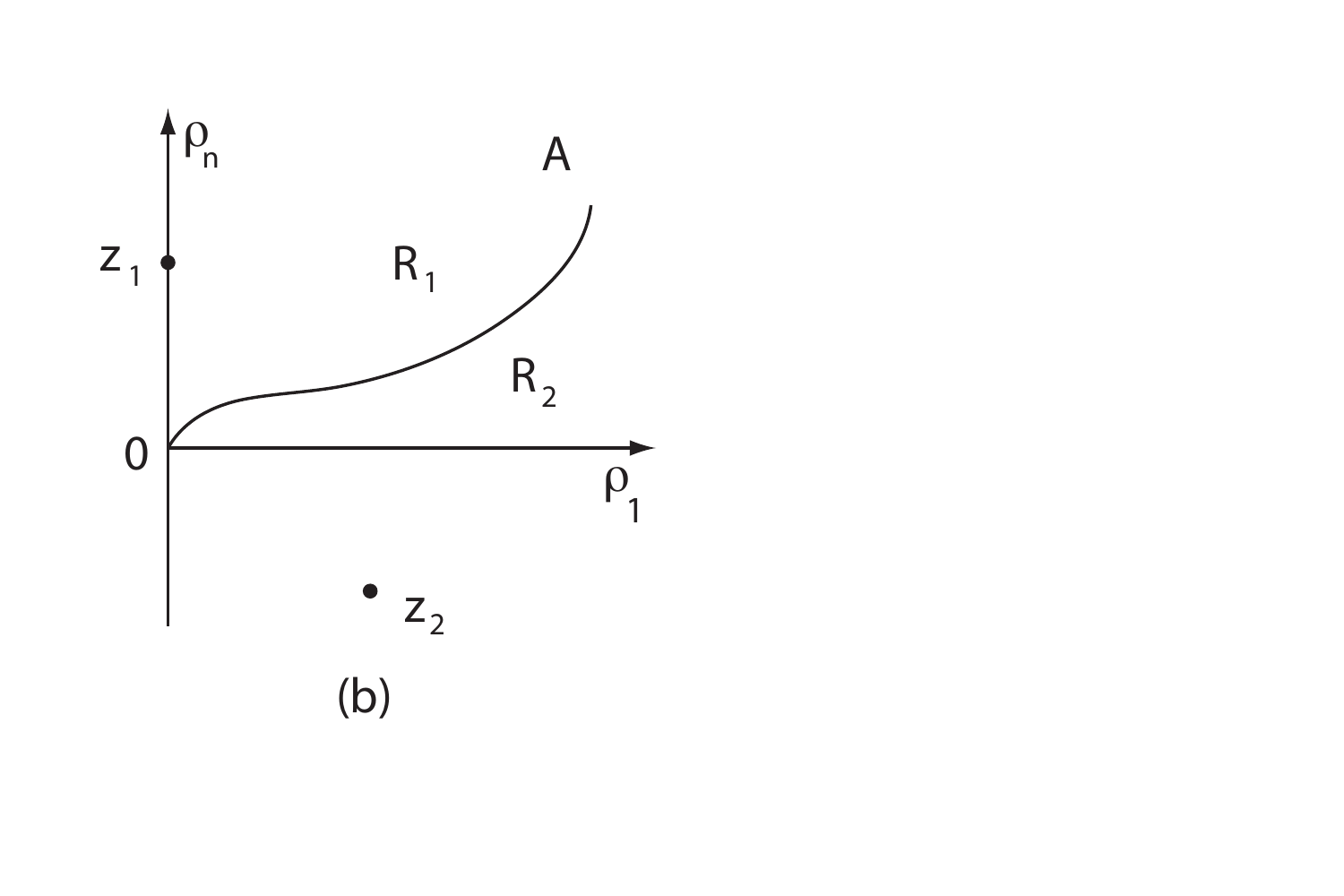}  
  \includegraphics[width=0.2\textwidth]{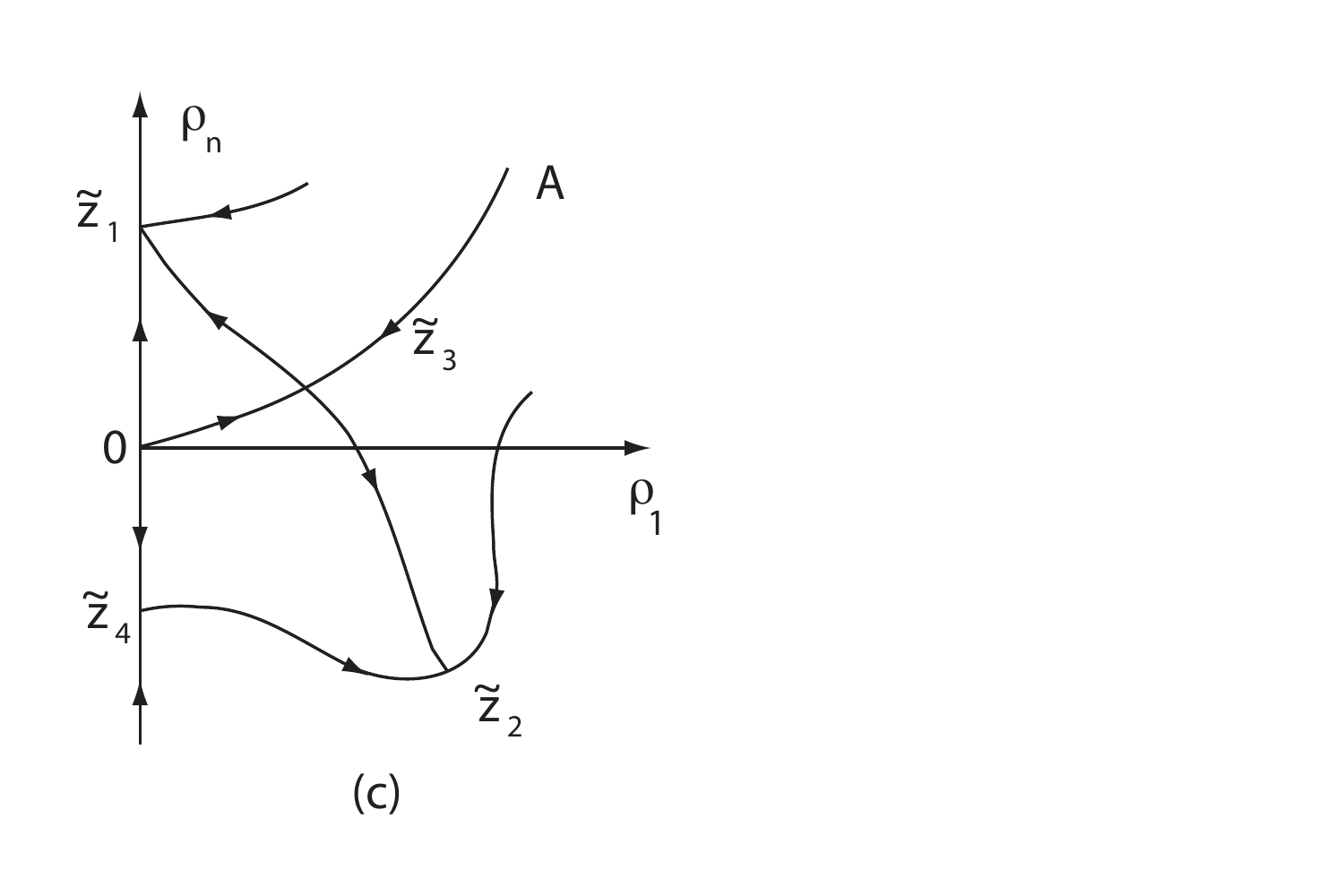}
  \includegraphics[width=0.2\textwidth]{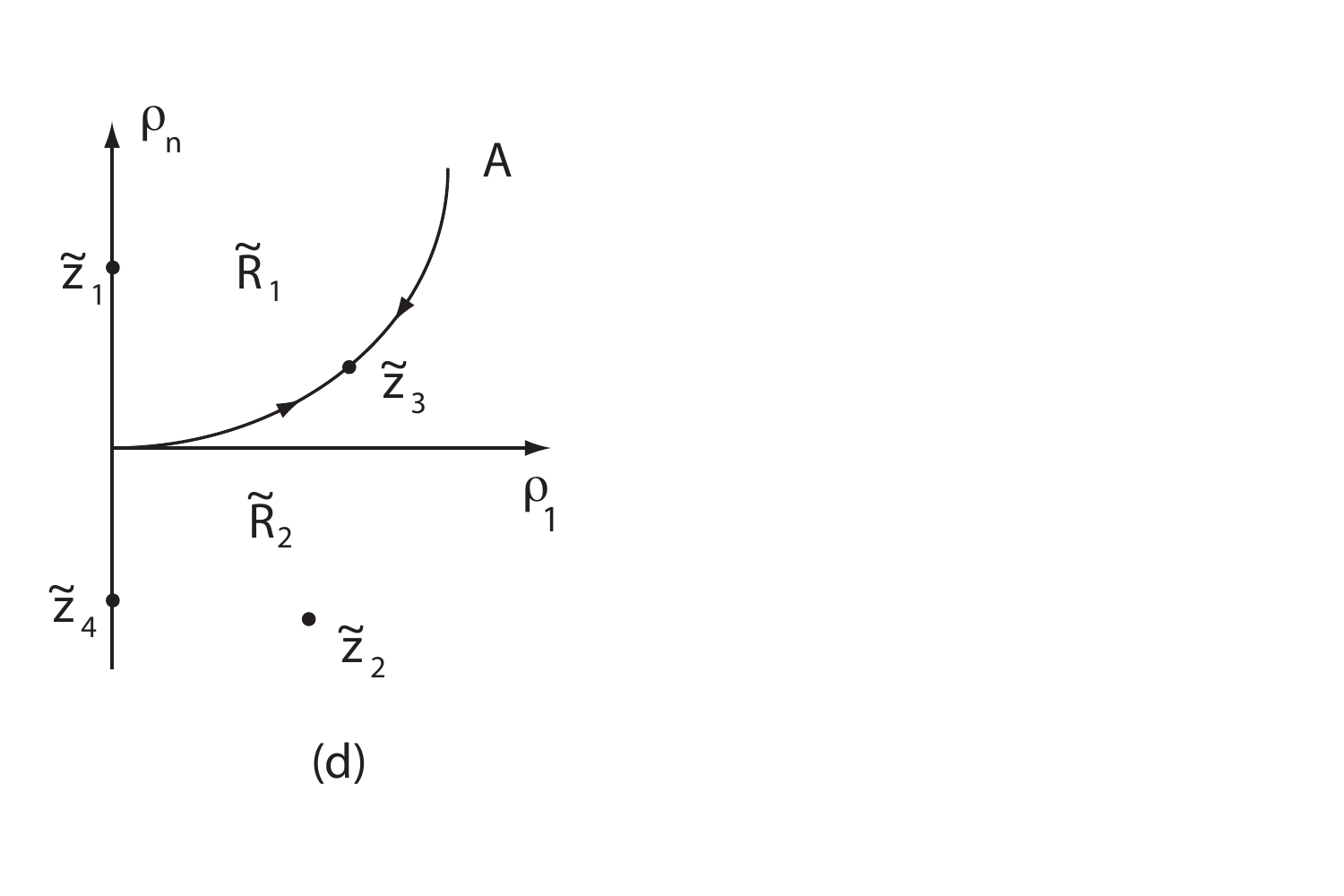}
  \caption{}\la{f8-37-1}
 \end{figure}

\subsection{First phase transition}
On the coexistence curve $\widehat{bc}$, 
$$\lambda_1=0, \lambda_2 <  0, \lambda_3 <0.$$
Hence the first phase transition crossing $\widehat{bc}$ is between  normal fluid state and  phase A superfluid state. 

On the coexistence curve $\widehat{cd}$, 
$$\lambda_1< 0, \lambda_2 =  0, \lambda_3 <0.$$
In this case, the first phase transition crossing this coexistence curve is between the normal fluid state and the phase C superfluid state.

\subsection{Second phase transitions}
When $(T,p)$ crosses the curve segment $\widehat{bcd}$,  (\ref{8.233}) will undergo a second transition. We need to consider two cases.

{\bf Second transition crossing $\widehat{f'c}$}

If $(T,p)$
passes through this curve segment $\widehat{bc}$, then the first transition solution is 
given by 
\begin{align*}
&(\rho_1,\rho_2,\rho_n)=(\rho^*_1,0,\rho^*_n),\\
&\rho^*_1=\rho^*_1(T,p)>0,\\
&\rho^*_n=\rho^*_n(T,p)<0.
\end{align*} 
Take a transformation
$$ \rho^{\prime}_1=\rho_1-\rho^*_1,\ \ \ \ \rho^{\prime}_2=\rho_2,\
\ \ \ \rho^{\prime}_n=\rho_n-\rho^*_n.$$ 
Then, the system
(\ref{8.233}) is in the following form (for simplicity, we drop
the primes):
\begin{equation}
\left.
\begin{aligned}
&\frac{d\rho_1}{dt}=\widetilde{\lambda}_1\rho_1-a_1\rho^*_1\rho_n-a_2\rho^*_1\rho_2-a_2\rho_1\rho_2-a_3\rho^3_1,\\
&\frac{d\rho_2}{dt}=\widetilde{\lambda}_2\rho_2-b_1\rho_n\rho_2-b_2\rho_1\rho_2-b_3\rho^2_2,\\
&\frac{d\rho_n}{dt}=\widetilde{\lambda}_3\rho_n-c_1\rho_1-c_2\rho_2+(c_3-3c_4\rho^*_n)\rho^2_n-c_4\rho^3_n,
\end{aligned}
\right.\label{8.234}
\end{equation}
where
\begin{equation}
\begin{aligned}
& \widetilde{\lambda}_1=\lambda_1+a_1|\rho^*_n|-2a_3\rho^*_1,\\
& \widetilde{\lambda}_2=\lambda_2+b_1|\rho^*_n|-b_2\rho^*_1,\\
& \widetilde{\lambda}_3=\lambda_3-2c_3|\rho^*_n|-3c_4\rho^{*2}_n.
\end{aligned}
\label{8.235}
\end{equation}
The linear operator in (\ref{8.234}) reads
$$L=
\left(\begin{matrix}
\widetilde{\lambda}_1    &     -a_2\rho^*_1           &      -a_1\rho^*_1\\
0                                    & \widetilde{\lambda}_2  &          0\\
-c_1                               & -c_2                             &  \widetilde{\lambda}_3
\end{matrix}
\right).$$ 
The three eigenvalues of $L$ are 
\begin{equation}
\beta_1=\widetilde{\lambda}_2,\quad 
\beta_{\pm}=\frac{1}{2}\left[\widetilde{\lambda}_1+\widetilde{\lambda}_3\pm\sqrt{(\widetilde{\lambda}_3-
\widetilde{\lambda}_1)^2+4a_1c_1\rho^*_1} \right].
\label{8.236}
\end{equation}
It is known that the transition solution $(\rho^*_1,0,\rho^*_n)$
is stable near $\widehat{bc}$. Therefore the eigenvalues of $L$
satisfy
$$
\beta_1(T,p)<0,\ \ \ \ \beta_{\pm}(T,p)<0\ \text{for}\ (T,p)\
\text{near}\ \widehat{bc}.
$$ 
However, near $\widehat{fc}$ there is
a curve segment $\widehat{f^{\prime}c}$ such that
$$\widehat{f^{\prime}c}=\{(T,p)\in \R^2_+|\ \beta_1(T,p)=0\}.$$
Thus, system (\ref{8.234}) has a transition on
$\widehat{f^{\prime}c}$, which is called the second transition of
(\ref{8.233}), and $\widehat{f^{\prime}c}$ is the coexistence
curve of phases $A$ and $B$; see Figure \ref{f8.41}.

\medskip

{\bf Second transition crossing $\widehat{ch'}$}

If $(T,p)$
passes through this curve segment $\widehat{bc}$, then the first transition solution is 
given by 
\begin{align*}
&(\rho_1,\rho_2,\rho_n)=(0, \eta_2,\eta_n),\\
&\eta_2=\eta_2(T,p)>0,\\
&\eta_n=\eta_n(T,p)<0.
\end{align*} 
Take a transformation
$$ \rho^{\prime}_1=\rho_,\ \ \ \ \rho^{\prime}_2=\rho_2-\eta_2,\
\ \ \ \rho^{\prime}_n=\rho_n-\eta_n.$$ 
Then, the system
(\ref{8.233}) is in the following form (for simplicity, we drop
the primes):
\begin{equation}
\left.
\begin{aligned}
&\frac{d\rho_1}{dt}=\widetilde{\widetilde{\lambda}}_1\rho_1-a_1\rho_1\rho_n-a_2\rho_1\rho_2 -a_3\rho^2_1,\\
&\frac{d\rho_2}{dt}=\widetilde{\widetilde{\lambda}}_2\rho_2-b_1\eta_2\rho_n
-b_2\eta_2\rho_1  - b_1 \rho_2 \rho_n  -b_2 \rho_1 \rho_2 -b_3\rho^2_2,\\
&\frac{d\rho_n}{dt}=\widetilde{\widetilde{\lambda}}_3\rho_n-c_1\rho_1-c_2\rho_2+(c_3-3c_4\eta_n)\rho^2_n-c_4\rho^3_n,
\end{aligned}
\right.\label{8.234-1}
\end{equation}
where
\begin{equation}
\begin{aligned}
& \widetilde{\widetilde{\lambda}}_1=\lambda_1+a_1|\eta_n|-a_2\eta_2,\\
& \widetilde{\widetilde{\lambda}}_2=\lambda_2+b_1|\eta_n|-2 b_2\eta_2,\\
& \widetilde{\widetilde{\lambda}}_3=\lambda_3-2c_3|\eta_n|-3c_4\eta^{2}_n.
\end{aligned}
\label{8.235-1}
\end{equation}
The linear operator in (\ref{8.234-1}) reads
$$L=
\left(\begin{matrix}
\widetilde{\widetilde{\lambda}}_1    &     0           &      -0 \\
-b_2 \eta_2                                  & \widetilde{\widetilde{\lambda}}_2  &          -b_1 \eta_2\\
-c_1                               & -c_2                             &  \widetilde{\widetilde{\lambda}}_3
\end{matrix}
\right).$$ 
The three eigenvalues of $L$ are 
\begin{equation}
\beta_1=\widetilde{\widetilde{\lambda}}_1,\quad 
\beta_{\pm}=\frac{1}{2}\left[\widetilde{\widetilde{\lambda}}_1+\widetilde{\widetilde{\lambda}}_3\pm\sqrt{( \widetilde{\widetilde{\lambda}}_3-
\widetilde{\widetilde{\lambda}}_1)^2+4b_1c_2\eta_2} \right].
\label{8.236-1}
\end{equation}
It is known that the transition solution $(\rho^*_1,0,\rho^*_n)$
is stable near $\widehat{bc}$. Therefore the eigenvalues of $L$
satisfy
$$
\beta_1(T,p)<0,\ \ \ \ \beta_{\pm}(T,p)<0\ \text{for}\ (T,p)\
\text{near}\ \widehat{cd}.
$$ 
However, near $\widehat{hc}$ there is
a curve segment $\widehat{h^{\prime}c}$ such that
$$\widehat{h^{\prime}c}=\{(T,p)\in \R^2_+|\ \beta_1(T,p)
=\widetilde{\widetilde{\lambda}}_1=0\}.$$
Thus, system (\ref{8.234-1}) has a transition on
$\widehat{h^{\prime}c}$, which is called the second transition of
(\ref{8.233}), and $\widehat{h^{\prime}c}$ is the coexistence
curve of phases $C$ and $B$; see Figure \ref{f8.41}.

\medskip

In summary, with the above analysis and the dynamic transition theory, we arrive at the following transition  theorem:

\bt\la{t8.14-2}
Define a few regions in  the  $PT$-plane (see
Figure \ref{f8.41}) by
\begin{align*}
&{E}_1=\{(T,p)\in \R^2_+\ |\ 
(\lambda_1,\lambda_2,  \lambda_3) (T,p)= (-, -, +)\},\\
&{E}_2=\{(T,p)\in \R^2_+\ |\ 
(\lambda_1,\lambda_2,  \lambda_3) (T,p)=(-, -, -) \},\\
&{H}_1=\{(T,p)\in \R^2_+\ |\ 
(\lambda_1,\widetilde{\lambda}_2,  \lambda_3) (T,p)= (+, -, +)\},\\
&{H}_2=\{(T,p)\in \R^2_+\ |\ 
(\lambda_1, \widetilde{\lambda}_2,  \lambda_3) (T,p)=(+, +, +)\},\\
& \text{Region }f'bc = 
  \{(T,p)\in \R^2_+\ |\ 
  (\lambda_1,\lambda_2, \widetilde{\lambda}_2, \lambda_3) (T,p)\\
        &    \qquad\qquad \qquad  \qquad \qquad =(+, -, -, -) \},\\
& \text{Region }cdh' = 
  \{(T,p)\in \R^2_+\ |\ 
  (\lambda_1,\widetilde{\widetilde{\lambda}}_2, \lambda_2,  \lambda_3) (T,p)\\
  & \qquad \qquad \qquad \qquad \qquad =
(-, -, +, -) \},\\
\end{align*}
and let the Region $0g'f'ch'$  be  the complement of the sum of the above regions.
Then the following conclusions hold true:

\begin{itemize}

\item[(1)] If $(T,p)\in {E}_1$,  the phase of $^3$He is
in solid state.

\item[(2)] If  $(T,p)\in E_2$,  the phase is in normal liquid state.

\item[(3)] If $(T, p) \in \text{Region } f'bc$, the phase is in phase $A$ superfluid state.

\item[(4)] If $(T, p) \in \text{Region } cdh'$, the phase is in phase $C$ superfluid state.

\item[(5)] If $(T, p) \in \text{Region } 0g'f'ch'$, the phase is in phase $B$ superfluid state.

\item[(6)] If $(T,p)\in {H}_1$,  there are two regions
$R_1$ and $R_2$ in the state space
$(\rho_1, \rho_2, \rho_n)$  such that, under a fluctuation which is described by the
initial value $(x_0,y_0, z_0)$ in (\ref{8.233}):  If
$(x_0,y_0, z_0)\in  {R}_1$ then the phase is in solid state,
and if $(x_0,y_0, z_0)\in  {R}_2$ then it is in phase $A$ superfluid
state.

\item[(7)] If $(T,p)\in {H}_2$,  there are two regions
$K_1$ and $K_2$ in the state space
$(\rho_1, \rho_2, \rho_n)$  such that  if
$(x_0,y_0, z_0)\in  K_1$ then the phase is in solid state,
and if $(x_0,y_0, z_0)\in {K}_2$ then it is in phase $B$ superfluid
state.

\end{itemize}
\et

\section{Classification of Superfluid Transitions}
In this section, we classify  the superfluid transitions of (\ref{8.233}) crossing various coexistence curves. 

First we consider the transitions crossing
curve segments $\widehat{cd}$ and $\widehat{bc}$ in Figure \ref{f8.41}. Obviously, we have  
\begin{align*}
& \widehat{cd}=\{(T,p)\in \R^2_+\ |\   (\lambda_1,  \lambda_2, \lambda_3) (T,p)=(-, 0, -)\}, \\
&
\widehat{bc} =\{(T,p)\in \R^2_+\ |\ (\lambda_1,  \lambda_2, \lambda_3) (T,p)=(0, -, -) \}.
\end{align*}
Let
\begin{align*}
& A_1=a_1c_1-a_3|\lambda_3|,  && A_2=a_1c_2-a_2|\lambda_3|,\\
&
B_1=b_1c_2-b_3|\lambda_3|, && B_2=b_1c_1-b_2|\lambda_3|.
\end{align*}
In fact, it is obvious that $A_2=B_2$.

\bt\la{t8.15}
For the system (\ref{8.233}) we have the following assertions:

\begin{itemize}

\item[(1)]  As $(T_0,p_0)\in \widehat{cd}$, the transition of
(\ref{8.233}) at $(T_0,p_0)$ is between the phase $B$ and normal
liquid. Furthermore if $B_1\leq 0$, then it is Type-I,   and if $B_1>0$, then it is
Type-II.

\item[(2)]  As $(T_0,p_0)\in\widehat{bc}$, the transition is between the phase $A$ and normal liquid. Moreover, if $A_1\leq 0$, then  it is Type-I,   and if $A_1>0$, then it is Type-II.

\end{itemize}
\et

Theorem \ref{t8.15} provides conditions  for the first transition of
(\ref{8.233}). The following theorem gives sufficient conditions
for the second transition near $\widehat{fc}$ in Figure \ref{f8.41}.

Obviously, the  curve $\widehat{fc}$  is given by 
$$\widehat{fc}=\{(T,p)\in \R^2_+\ |\   (\lambda_1,  \lambda_2,  \lambda_3)(T,p) 
=(+, 0, -)\}.
$$ 

To set up the second transition theorem, we need to assume the following conditions.
Let $\varepsilon >0$ be small. Suppose that
\begin{equation}
B_2>-\varepsilon ,\quad   b_1a_3-b_2a_1> -\varepsilon  \quad 
\text{for}\ (T,p)\in\widehat{fc},\label{8.237}
\end{equation}
and the gap between $\widehat{bc}$ and $\widehat{fc}$ is small, i.e.,
\begin{equation}
|T_2-T_3|=O(\varepsilon )\quad    \forall (T_2,p)\in\widehat{bc},\ 
\ (T_3,p)\in\widehat{fc}.\label{8.238}
\end{equation}
We also assume that
\begin{align}
& (a_1,c_1)=O(\varepsilon ),\ \ \ \  ( b_1,c_2,c_3,a_3, b_3)=O(1),\label{8.239} \\
& a_3B_1-b_2A_2>0 \qquad   \text{in}\ \widehat{fc}\ \text{with}\ A_1\leq
0.\label{8.240}
\end{align}

\bt\la{t8.16}
Under conditions (\ref{8.237}) and
(\ref{8.238}), there exists a curve segment
$\widehat{f^{\prime}c}$ near
$\widehat{fc}$ as shown in Figure \ref{f8.41} such that
(\ref{8.233}) has the second transition from the first transition
solution $(\rho^*_1,0,\rho^*_n)$, i.e., (\ref{8.234}) has a
transition from $(\rho_1,\rho_2,\rho_n)=0$ in $\widehat{f'c}$,
and the transition solutions
$(\widetilde{\rho}_1,\widetilde{\rho}_2,\widetilde{\rho}_n)$
satisfy that $\widetilde{\rho}_2>0$. In addition, if (\ref{8.239})
and (\ref{8.240}) hold  true, then this transition is Type-II.
\et

Physical experiments display that the superfluid transition of
liquid $^3$He between the normal liquid and superfluid phase $B$
is continuous. Hence, it is necessary to give the conditions of
Type-I transition of (\ref{8.233}) at the intersecting point $C$
of two curves $\lambda_1=0$ and $\lambda_2=0$.

\bt\la{t8.17} 
Let $(T_0,p_0)$ be the point $C$ that
$\lambda_1(T_0,p_0)=0$ and $\lambda_2(T_0,p_0)=0$. Then the
transition of (\ref{8.233}) at $(T_0,p_0)$ is Type-I if and only
if one of the following two conditions hold true:

\begin{itemize}

\item[(i)]  $A_1\leq 0, B_1\leq 0, A_2=B_2<0$,

\item[(ii)] $A_1\leq 0, B_1\leq 0, A_2=B_2\geq 0$ and
$A_1B_1>A_2B_2$.
\end{itemize}
In particular, if the transition is Type-I, then for
$\lambda_1>0, \lambda_2>0$ near $(p_0,T_0)$, there are four types
of topological structure of the transition on center manifold, which
are classified as follows:

\begin{itemize}

\item[(1)] This transition is of the structure as shown in Figure
\ref{f8.42}(a),  if  
$$\lambda_1|B_1|+\lambda_2A_2>0\quad \text{and} \quad 
\lambda_2|A_1|+\lambda_1B_2>0.$$

\item[(2)] The transition is of the structure as shown in Figure
\ref{f8.42}(b), if  
$$\lambda_1|B_1|+\lambda_2A_2<0 \quad \text{and} \quad 
\lambda_2|A_1|+\lambda_1B_2<0.$$

\item[(3)] The transitions is of the structure as shown in Figure
\ref{f8.42}(c), if  
$$\lambda_1|B_1|+\lambda_2A_2<0 \quad \text{and} \quad
\lambda_2|A_1|+\lambda_1B_2>0.$$

\item[(4)] The transition has the structure as shown in Figure
\ref{f8.42}(d),  if 
$$\lambda_1|B_1|+\lambda_2A_2>0\quad \text{and} \quad
\lambda_2|A_1|+\lambda_1B_2<0.$$
\end{itemize}
\et

\begin{figure}[hbt]
  \centering
  \includegraphics[width=0.2\textwidth]{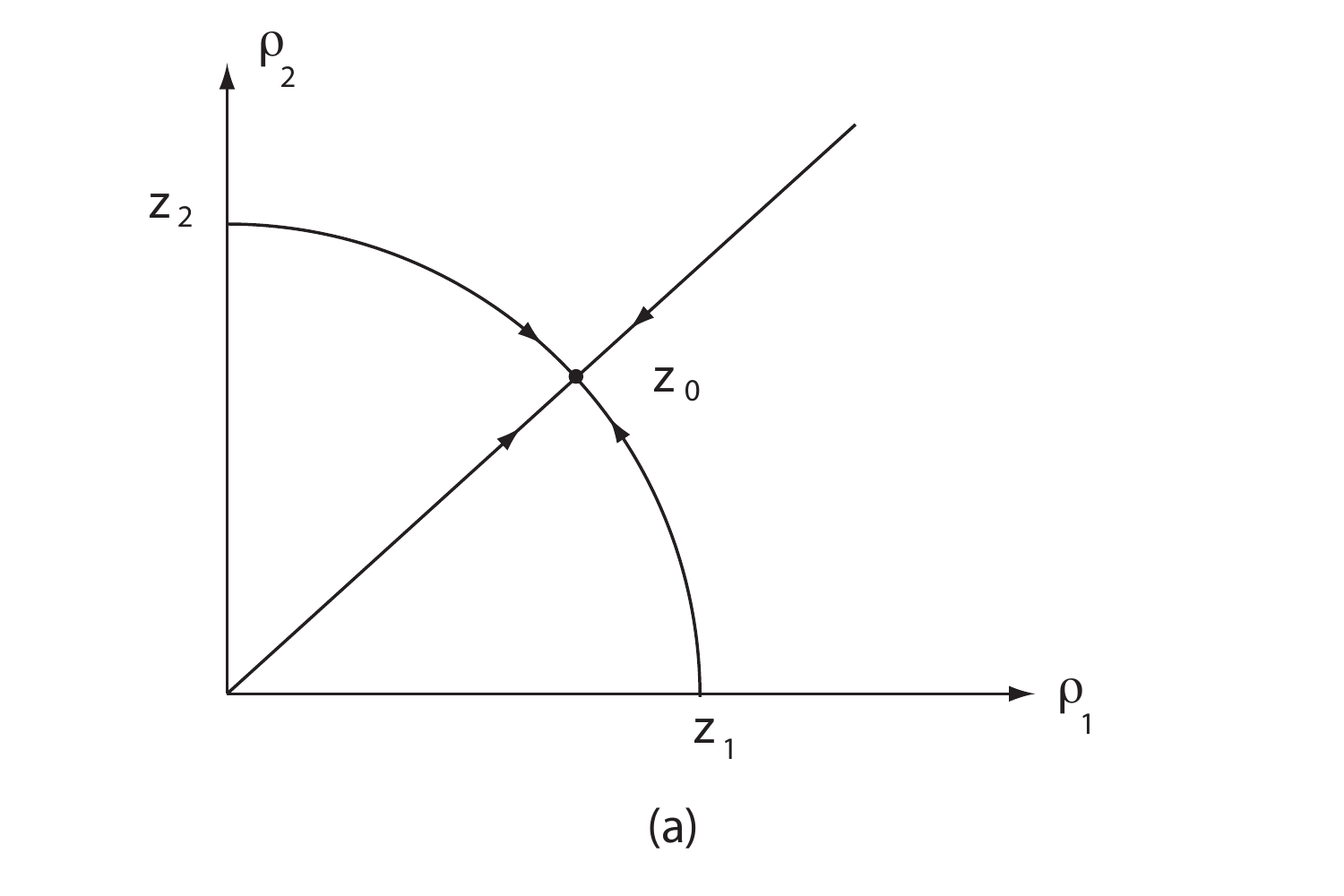}
  \includegraphics[width=0.2\textwidth]{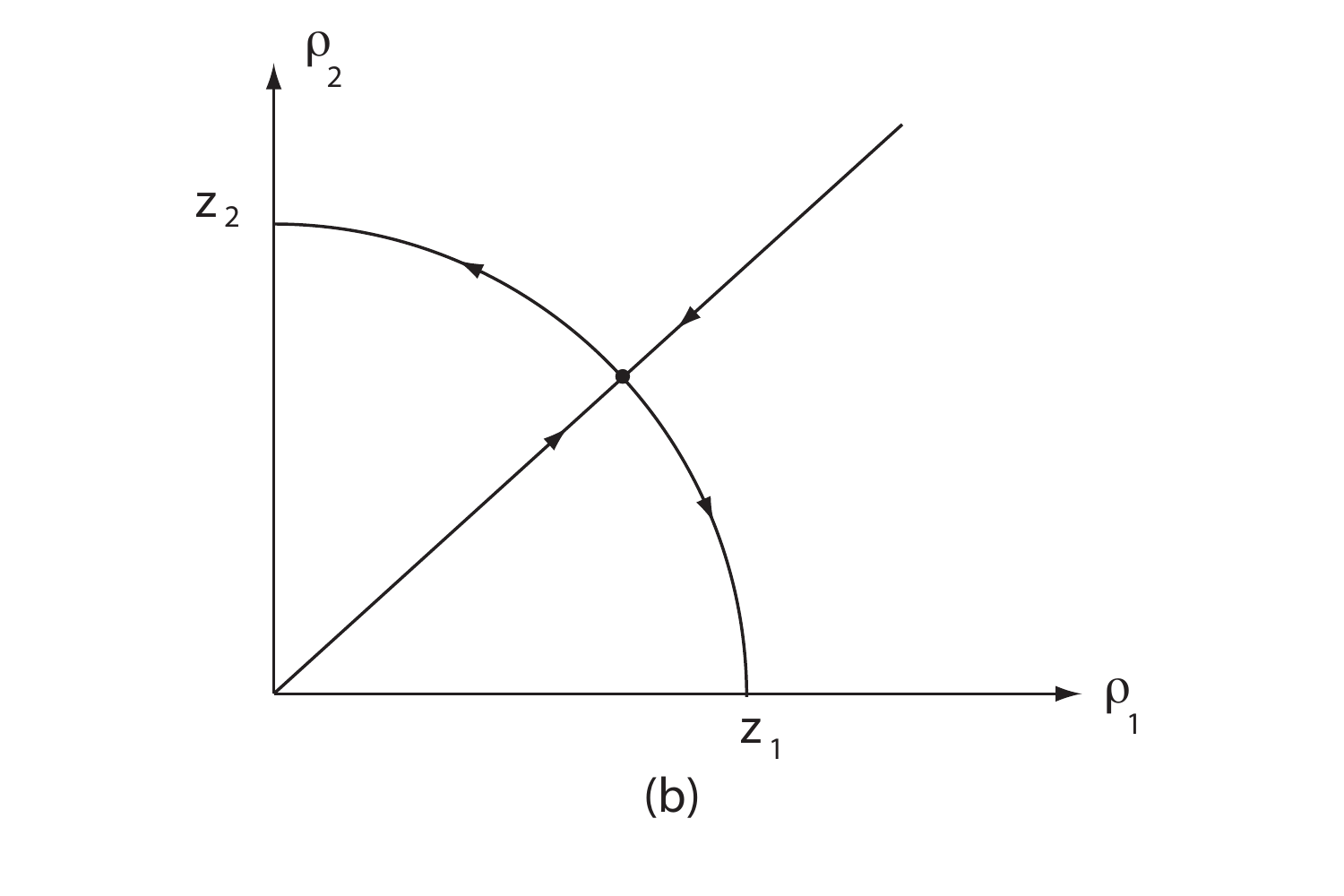}
  \includegraphics[width=0.2\textwidth]{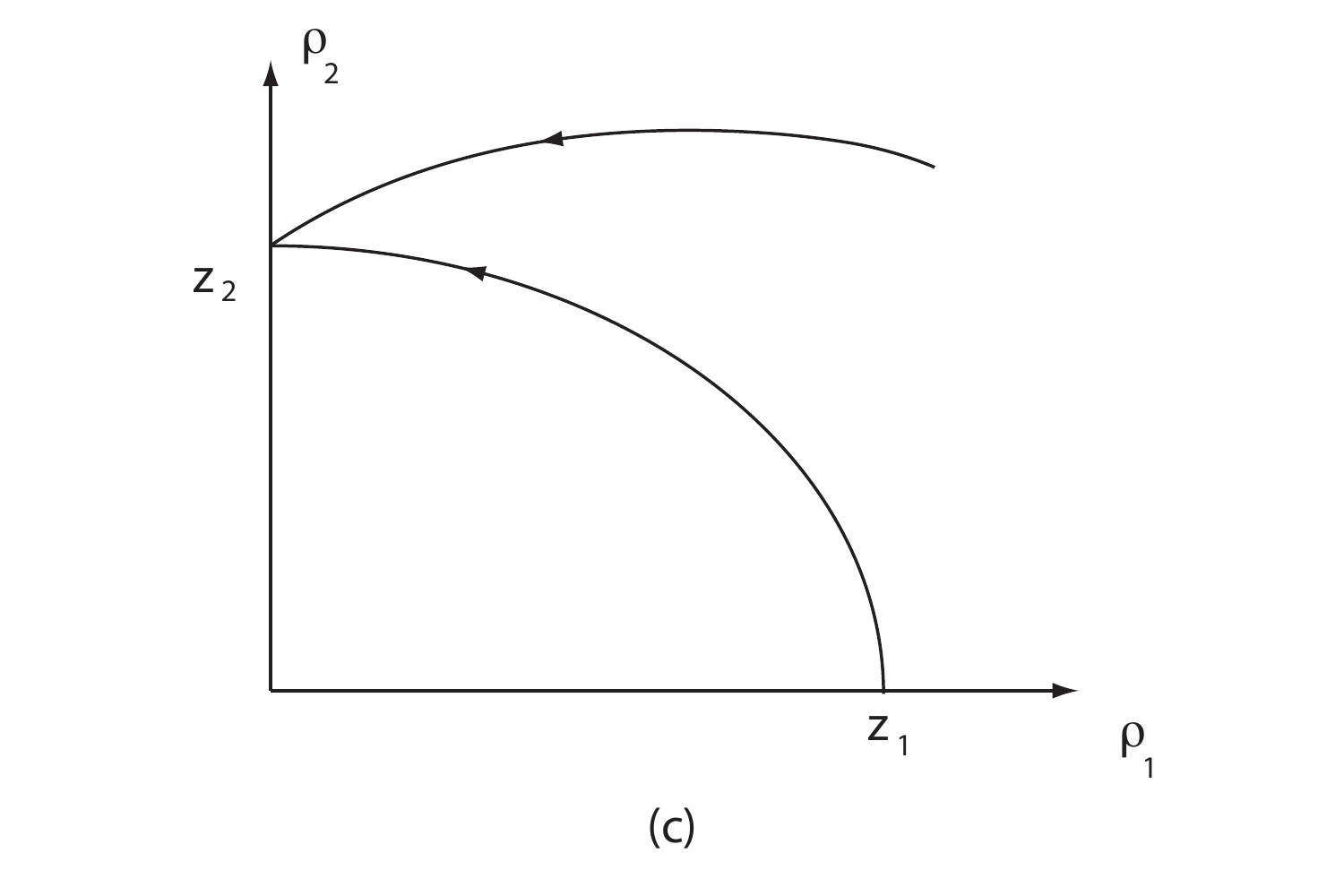}
  \includegraphics[width=0.2\textwidth]{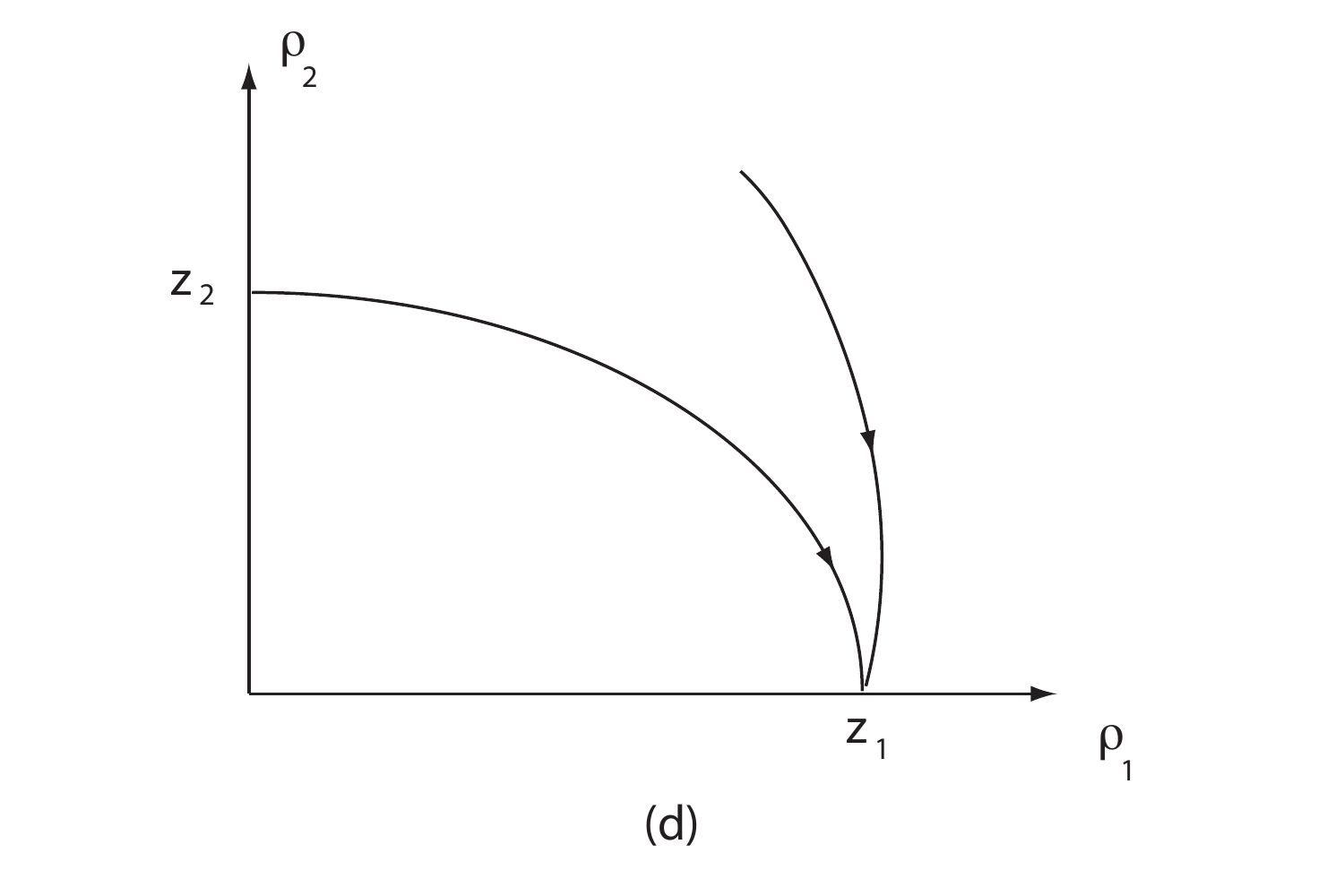}
  \caption{Topological structure of  Type-I transition
near the intersection point $C=(T_0,p_0)$ of $\lambda_1=0$ and
$\lambda_2=0$.}\la{f8.42}
 \end{figure}
Before  the proof of these theorems., we have the following remark.

\br\la{rhe3}
{\rm
Physically, the transition between normal liquid and superfluid
for $^3$He is generally Type-I, the transition between
superfluid phases $A$ and $B$ is Type-II, and the region of phase
$A$ is narrow. Therefore, under the conditions
(\ref{8.237})-(\ref{8.240}) and
\begin{align*}
& B_1<0  &&\text{in}\ \widehat{cd},\\
& B_1\leq 0, A_1<0 &&  \text{near point}\
C=\bar{\Gamma}_1\cap\bar{\Gamma}_2 = (T_0, p_0),
\end{align*}
the above theorems (Theorems \ref{t8.15}-\ref{t8.17}) provide a precise
mathematical proof for superfluid transitions of liquid $^3$He 
with no applied magnetic field.

By condition (\ref{8.239}) we see that
$$a_1\cdot c_1=O(\varepsilon^2),\ \ \ \ a_3=O(1).$$
Assertion (1) of Theorem \ref{t8.15}  implies that only in a
very small range of $(T,p)$ near the change point of superfluid
and solid, the transition between normal liquid and superfluid is
II-type, and this range is
$$0>\lambda_3(T,p)>-\frac{a_1c_1}{a_3}=-O(\varepsilon^2).$$
Moreover, the superfluid density of phase $A$ near the solid phase
is in the quantitative order $\varepsilon^3$, i.e.,
$$\rho_1=\frac{a^2_1c_1}{a^2_3c_3}=O(\varepsilon^3).$$
Hence, difference between the Type-I and  the Type-II
phase transitions in experiments is very small.

Note  that Theorem \ref{t8.16} are also valid if 
condition (\ref{8.237}) is replaced by that
$$\frac{\partial\lambda_2}{\partial T}\gg 1\ \ \ \ \text{for}\
(T,p)\ \text{near}\ \widehat{fc}.$$ 
}
\er


\bp[Proof of Theorem \ref{t8.15}] As $(T_0,p_0)\in\widehat{bc},
\lambda_2(T_0,p_0)<0$ and the space of $(\rho_1,0,\rho_n)$ is
invariant for (\ref{8.233}). Therefore, the transition equations
of (\ref{8.233}) at $(T_0,p_0)$ are referred to the following form
\begin{equation}
\left.
\begin{aligned}
&\frac{d\rho_1}{dt}=\lambda_1\rho_1-a_1\rho_n\rho_1-a_3\rho^2_1,\\
&\frac{d\rho_n}{dt}=\lambda_3\rho_n-c_1\rho_1+c_3\rho^2_n-c_4\rho^3_n.
\end{aligned}
\right.\label{8.241}
\end{equation}

The second order approximation of  the center manifold function
$\rho_n$ of (\ref{8.241}) satisfies the equation
$$\lambda_3\rho_n+c_3\rho^2_n=c_1\rho_1.$$
Its solution is
$$\rho_n=-\frac{c_1\rho_1}{|\lambda_3|}+\frac{c^2_1c_3}{|\lambda_3|^3}\rho^2_1+o(\rho^2_1).$$
Putting $\rho_n$ in the first equation of (\ref{8.241}) we get the
reduced equation of (\ref{8.233}) on center manifold as follows
\begin{equation}
\frac{d\rho_1}{dt}=\lambda_1\rho_1+\frac{1}{|\lambda_3|}A_1\rho^2_1-\frac{a_1c_1c_3}{|\lambda_3|^3}\rho^3_1+o(\rho^3_1).\label{8.242}
\end{equation}
Assertion (2) follows from (\ref{8.242}).

Likewise, if $(T_0,p_0)\in\widehat{cd}$,   $\lambda_1(T_0,p_0)<0$ and the
space of $(0,\rho_2,\rho_n)$ is invariant for (\ref{8.233}),
therefore in the same fashion as above we can prove Assertion (1).
The proof is complete.
\ep

\bp[Proof of Theorem \ref{t8.16}] 
We proceed in the following two
cases.

{\sc Case 1:  
 $A_1\leq 0$ in $\widehat{bc}$}. In this case, by Theorem
\ref{t8.15}, the transition of (\ref{8.233}) in $\widehat{bc}$ is Type-I,
and the transition solution $(\rho^*_1,0,\rho^*_n)$ satisfies that
$$\rho^*_n=-\frac{c_1}{|\lambda_3|}\rho^*_1.$$
The equations describing the second transition are given by
(\ref{8.234}) and the eigenvalues in (\ref{8.235}) and
(\ref{8.236}) are rewritten as
\begin{align}
\beta_1=  & \lambda_2+b_1|\rho^*_n|-b_2\rho^*_1\label{8.243}\\
=&\lambda_2+\frac{1}{|\lambda_3|}B_2\rho^*_1,\nonumber\\
\beta_+=&\frac{1}{2}[\widetilde{\lambda}_1+\widetilde{\lambda}_3+\sqrt{(\widetilde{\lambda}_1-
\widetilde{\lambda}_3)^2+4a_1c_1\rho^*_1}]\label{8.244}\\
=&(\text{by}\ (\ref{8.238})\ \text{and}\ \rho^*_1,
\widetilde{\lambda}_1\ \text{being  small})\nonumber\\
=&-\lambda_1+o(|\lambda_1|),\nonumber\\
\beta_1<  &\beta_+.\label{8.245}
\end{align}
In addition, we know that
\begin{align*}
& \lambda_1=0,\ \ \ \ \lambda_2<0  &&  \text{on}\ \widehat{bc},\\
& \lambda_2(T-\delta ,p)>0  &&  \text{for}\ (T,p)\in\widehat{fc}\ \text{and}\
\delta >0.
\end{align*}
Hence, by assumptions (\ref{8.237}) and (\ref{8.238}),
from (\ref{8.243})-(\ref{8.245}) we can infer that there exists a
curve segment $\widehat{f'c}$ near $\widehat{fc}$ such that for
$(T_2,p)\in\widehat{bc}$ and $(T_0,p)\in\Gamma^{\prime}_3$ we have
\begin{equation}
\beta_1(T,p)\left\{
\begin{aligned} 
& <0  &&  \text{ if } T_0<T\leq T_2,\\
& =0  &&   \text{ if } T=T_0,\\
& >0 &&  \text{ if } T<T_0,
\end{aligned}
\right.\label{8.246}
\end{equation}
and $\beta_-<\beta_+=-\lambda_1+o(|\lambda_1|)<0$. Hence, by
Theorem \ref{t5.1}, the system (\ref{8.234}) has a transition on
$\Gamma^{\prime}_3$.

To determine the transition type, we consider the center manifold
function of (\ref{8.234}), which     
satisfies
\begin{equation}
\left(\begin{matrix} 
\widetilde{\lambda}_1&-a_1\rho^*_1\\
-c_1&\widetilde{\lambda}_3
\end{matrix}
\right)\left(\begin{array}{l} \rho_1\\
\rho_n
\end{array}
\right)=\left(\begin{matrix} a_2\rho^*_1\rho_2\\
c_2\rho_2
\end{matrix}
\right)+o(\rho_2).\label{8.247}
\end{equation}
The solution of (\ref{8.247}) is
\begin{eqnarray*}
&&\rho_1=\frac{a_2\widetilde{\lambda}_3+c_2a_1}{\widetilde{\lambda}_3\widetilde{\lambda}_1-c_1a_1\rho^*_1}\rho^*_1\rho_2+o(\rho_2),\\
&&\rho_n=\frac{\widetilde{\lambda}_1c_2+c_1a_2\rho^*_1}{\widetilde{\lambda}_3\widetilde{\lambda}_1-c_1a_1\rho^*_1}\rho_2+o(\rho_2).
\end{eqnarray*}
From (\ref{8.235}), (\ref{8.238}) and (\ref{8.239}) we can obtain
\begin{equation}
\rho_1\simeq\frac{A_2\rho_2}{a_3|\lambda_3|},\ \ \ \ \rho_n\simeq
-\frac{c_2}{|\lambda_3|}\rho_2.\label{8.248}
\end{equation}
Inserting the center manifold function (\ref{8.248}) into the
second equation of (\ref{8.234}), we get the reduced equation as
$$\frac{d\rho_2}{dt}=\widetilde{\lambda}_2\rho_2+\frac{1}{|\lambda_3|}(B_1-\frac{b_2A_2}{a_3})\rho^2_2$$
By (\ref{8.240}),   
the transition of (\ref{8.234}) is Type-II.

\bigskip
{\sc  Case 2.}
 $A_1>0$ in $\widehat{bc}$. In this case, the transition of
(\ref{8.233}) in $\widehat{bc}$ is Type-II, and the transition
solution in $\widehat{bc}$ is
$$\rho^*_1=\frac{a_1}{a_3}|\rho^*_n|,\ \ \ \
\rho^*_n=-\frac{c_3}{2c_4}\left(\sqrt{1+\frac{4c_4c_1a_1}{a_3c^2_3}}-1\right).$$
The eigenvalue $\beta_1$ in (\ref{8.236}) reads
$$\beta_1=\lambda_2+\frac{1}{a_3}(b_1a_3-b_2a_1)|\rho^*_n|.$$
By (\ref{8.237}) and (\ref{8.238}) it implies that (\ref{8.246})
holds. Hence (\ref{8.233}) has a second transition in
$\Gamma^{\prime}_3$ for $A_1>0$ in $\widehat{bc}$.

Under the condition (\ref{8.239}), we have
$$\rho^*_n\simeq -\frac{a_1c_1}{a_3c_3},\ \ \ \
\widetilde{\lambda}_1\simeq -a_3\rho^*_1,\ \ \ \
\widetilde{\lambda}_3\simeq\lambda_3-\frac{2a_1c_1}{a_3}.$$ By
$A_1>0$ we get that $|\lambda_3|\leq 0(\varepsilon^2)$. Thus, the
solutions of (\ref{8.247}) can be rewritten as
\begin{eqnarray*}
&&\rho_1\simeq\frac{a_1c_1}{\widetilde{\lambda}_1\widetilde{\lambda}_3-c_1a_1\rho^*_1}\rho^*_1\rho_2\simeq\frac{c_2}{c_1}\rho_2,\\
&&\rho_2\simeq -\frac{a_3c_2}{a_1c_1}\rho_2.
\end{eqnarray*}
Putting $\rho_1$ and $\rho_n$ into the second equation of
(\ref{8.234}), we obtain reduced equation on the center manifold
as
$$\frac{d\rho_2}{dt}=\widetilde{\lambda}_2\rho_2+\left(\frac{b_1c_2a_3}{a_1c_1}-\frac{c_2b_2}{c_1}-b_3\right)\rho^2_2.$$
Due to (\ref{8.239}) we see that
$$\frac{b_1c_2a_3}{a_1c_1}-\frac{c_2b_2}{c_1}-b_3>0.$$
Therefore, this transition of (\ref{8.234}) is Type-II.

It is clear that the second transition solutions
$(\widetilde{\rho}_1,\widetilde{\rho}_2,\widetilde{\rho}_n)$
satisfy that $\widetilde{\rho}_2>0$. Thus, the theorem is
proved.
\ep

\bp[Proof of Theorem~\ref{t8.17}]  
At point $C=(T_0,p_0),
\lambda_1(T_0,p_0)=0, \lambda_2(T_0,p_0)=0$. Hence, the center
manifold function of (\ref{8.233}) at $(T_0,p_0)$ reads
$$\rho_n=-\frac{c_1}{|\lambda_3|}\rho_1-\frac{c_2}{|\lambda_3|}\rho_2+\frac{c_3}{|\lambda_3|^3}(c_1\rho_1+c_2\rho_2)^2.$$
Putting $\rho_n$ in the first two equations of (\ref{8.233}) we
get the reduced equations on the center manifold as 
\begin{equation}
\begin{aligned}
\frac{d\rho_1}{dt}=&  \lambda_1\rho_1+\frac{1}{|\lambda_3|}(A_1\rho^2_1+A_2\rho_1\rho_2)\\
& -\frac{c_3a_1}{|\lambda_3|^3}(c_1\rho_1+
c_2\rho_2)^2\rho_1, \\
\frac{d\rho_2}{dt}=& \lambda_2\rho_2+\frac{1}{|\lambda_3|}(B_1\rho^2_2+B_2\rho_1\rho_2)\\
& -\frac{c_3b_1}{|\lambda_3|^3}
(c_1\rho_1+c_2\rho_2)^2\rho_2.
\end{aligned}
\label{8.249}
\end{equation}
To verify the Type-I transition, by the attractor bifurcation
theorem \cite{b-book}, it suffices to consider the following
equations:
\begin{equation}
\begin{aligned}
&\frac{d\rho_1}{dt}=A_1\rho^2_1+A_2\rho_1\rho_2-\frac{c_3a_1}{|\lambda_3|^2}(c_1\rho_1+c_2\rho_2)^2\rho_1,\\
&\frac{d\rho_2}{dt}=B_1\rho^2_2+B_2\rho_1\rho_2-\frac{c_3b_1}{|\lambda_3|^2}(c_1\rho_1+c_2\rho_2)^2\rho_2.
\end{aligned}
\label{8.250}
\end{equation}

Since (\ref{8.230}) have variational structure, the flows of
(\ref{8.250}) are of gradient type. Therefore, $(\rho_1,\rho_2)=0$
has no elliptic region for (\ref{8.250}). Hence, in the same
fashion as used in Section 6.3 in \cite{b-book}, one can prove that the region
$$S=\{(\rho_1,\rho_2)\in \R^2|\ \rho_1>0, \rho_2>0\}$$
is a stable parabolic region. Namely, $(\rho_1,\rho_2)=0$ is an
asymptotically stable singular point of (\ref{8.250})  if and only
if one of the two conditions (i) and (ii) holds true. Thus, we
only need to prove Assertions (1)-(4).

For Type-I transition at point $C=(T_0,p_0)$, by condition (i) and
(ii), $A_1<0$ and $B_1<0$. Hence, as $\lambda_1>0, \lambda_2>0$
there are bifurcated solutions of (\ref{8.249}) in the
$\rho_1$-axis and $\rho_2$-axis as
$$z_1=(\rho^*_1,0)=\left(\frac{|\lambda_3|}{|A_1|}\lambda_1,0\right),\quad z_2=(0,\rho^*_2)=\left(0,\frac{|\lambda_3|}{|B_1|}\lambda_2\right).$$
The Jacobian matrices of (\ref{8.249}) at $z_1$ and $z_2$ are
given by
\begin{align*}
& J(z_1)=\left(\begin{matrix}
-\lambda_1&  * \\
0&\lambda_2+\frac{B_2}{|A_1|}\lambda_1
\end{matrix}
\right),\\
& 
J(z_2)=\left(\begin{matrix}
\lambda_1+\frac{A_2}{|B_1|}\lambda_2&0\\
*&-\lambda_2 \end{matrix} \right).
\end{align*} 
Since (\ref{8.249}) has at
most three bifurcated singular points in region $\bar{S}$, there
are only four types of  Type-I transitions, as shown in 
Figure \ref{f8.42}(a)-(d), and each type is completely determined by the signs
of the eigenvalues of $J(z_1)$ and $J(z_2)$. Thus, from the
eigenvalues $\beta
(z_1)=\frac{1}{|A_1|}(\lambda_2|A_1|+\lambda_1B_2)$ of $J(z_1)$
and $\beta (z_2)=\frac{1}{|B_1|}(\lambda_1|B_1|+\lambda_2A_2)$ of
$J(z_2)$ it is readily to derive Assertions (1)-(4). The proof is
complete.
\ep

\section{Liquid $^3$He with Nonzero Applied Field}

When liquid $^3$He is placed in a magnetic field $H$, the
superfluid transition is different from that with $H=0$.
Experiments show  that as a magnetic field is applied, a new
superfluid phase $A_1$ appears, and the region of phase $A$ can be
extended to the bottom at $p=0$. The $PT$-phase diagram is
schematically illustrated by Figure \ref{f8.43}.
\begin{figure}[hb]
  \centering
  \includegraphics[width=0.2\textwidth]{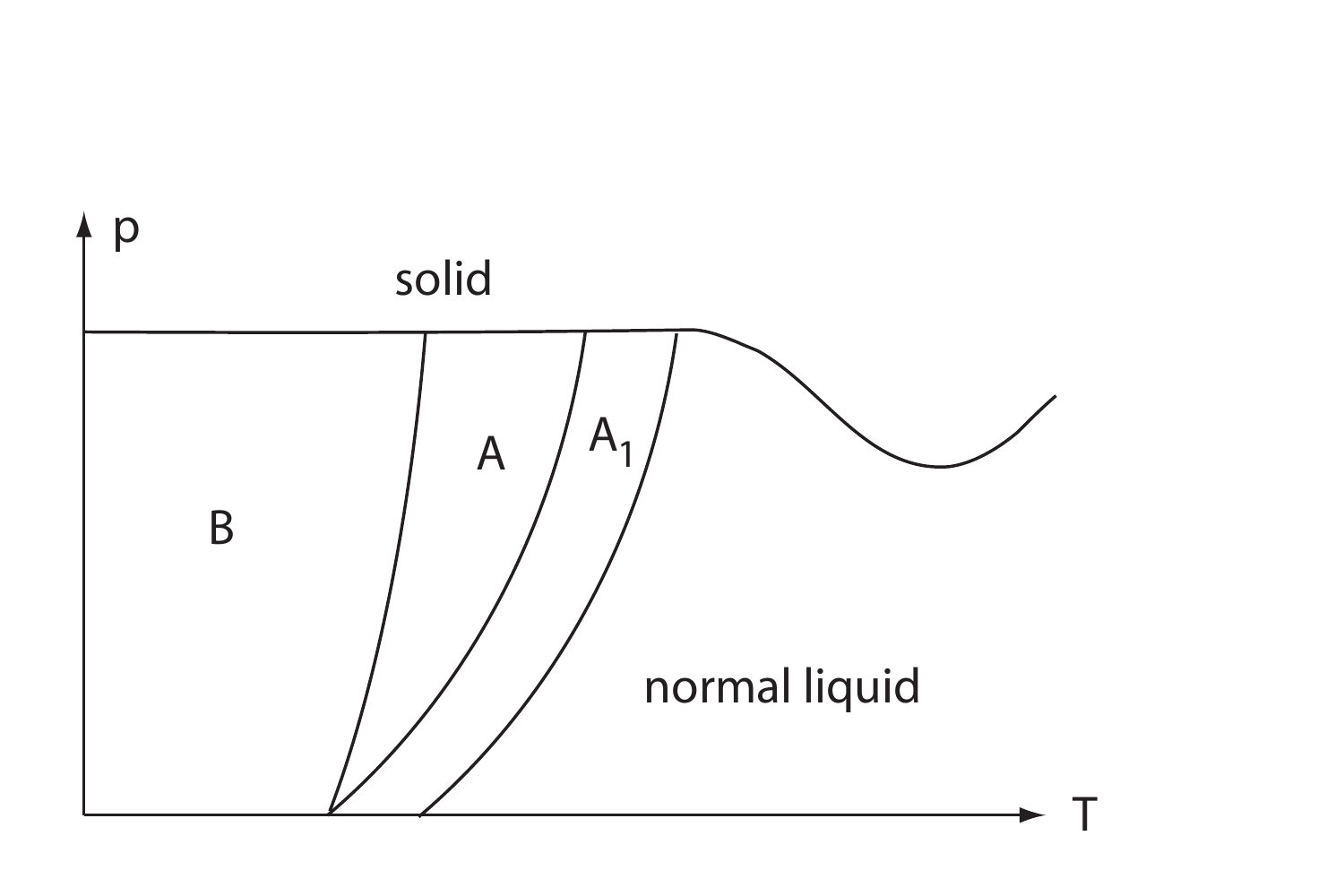}
  \caption{$PT$-phase diagram of $^3$He in a magnetic
field.}\la{f8.43}
 \end{figure}

As a magnetic field $H$ is applied, there is a new pairing state
to appear, in which the spin of pairing atoms is parallel to the
magnetic field. This new state corresponds to the phase $A_1$, and
is expressed by
$$\sqrt{2}|\Phi >=2a|\uparrow\uparrow >.$$

We introduce the complex valued functions $\psi_0$ to the state
$|\uparrow\uparrow >, \psi_1$ to the state $|\downarrow\downarrow
>$, and $\psi_2$ to this state $|\uparrow\downarrow
>+|\downarrow\uparrow >$. Let $\rho_0,\rho_a,\rho_b$ stand for the
densities of superfluid phases $A_1,A,B$ respectively. Then we
have
\begin{align*}
&\rho_0=|\psi_0|^2,\\
&\rho_a=\tau_0|\psi_0|^2+\tau_1|\psi_1|^2 &&  (\tau_0>0,
\tau_1>0),\\
&\rho_b=\tau_2|\psi_0|^2+\tau_3|\psi_1|^2+\tau_4|\psi_2|^2  && (\tau_2,\tau_3\geq 0, \tau_4>0),
\end{align*}
and the total density of liquid $^3$He in a magnetic field is
given by 
$$\rho =\left\{\begin{aligned} 
& \rho_n+\rho_0 &&\text{at  state}\ A_1,\\
& \rho_n+\rho_{\alpha} &&\text{at  state}\ A,\\
& \rho_n+\rho_b &&\text{at state}\ B.
\end{aligned}
\right.$$

Thus, similar to (\ref{8.223}), for liquid $^3$He with $H\neq 0$
we give the Ginzburg-Landau free energy in the following form. For
simplicity we take the nondimensional form:
\begin{widetext}
\begin{align}
G(\psi_0,\psi_1,\psi_2,\rho_n)
= & \frac{1}{2}\int_{\Omega}\Big[\kappa_0|\nabla\psi_0|^2-
\lambda_0|\psi_0|^2+\alpha_0\rho_n|\psi_0|^2+\alpha_1|\psi_0|^2|\psi_1|^2+\alpha_2|\psi^2_0||\psi_2|^2
     +\frac{\alpha_3}{2}|\psi_0|^4   \nonumber \\
& + \kappa_1|\nabla\psi_1|^2-\lambda_1|\psi_1|^2 
 +a_1\rho_n|\psi_1|^2+a_2|\psi_1|^2|\psi_2|^2+\frac{a_3}{2}|\psi_1|^4  \nonumber \\
& +\kappa_2|\nabla\psi_2|^2- \lambda_2|\psi_2|^2
 +b_1\rho_1|\psi_2|^2+\frac{b_3}{2}|\psi_2|^4
  +\kappa_3|\nabla\rho_n|^2-\lambda_3|\rho_n|^2-
\frac{c_3}{3}\rho^3_n-\frac{c_4}{4}\rho^4_n\Big]dx. \label{8.251}
\end{align}
The equations describing liquid $^3$He with $H\neq 0$ read
\begin{equation}
\left.
\begin{aligned} 
&\frac{\partial\psi_0}{\partial
t}=\kappa_0\Delta\psi_0+\lambda_0\psi_0-\alpha_0\rho_n\psi_0-\alpha_1|\psi_1|^2\psi_0-\alpha_2|
\psi_2|^2\psi_0-\alpha_3|\psi_0|^2\psi_0,\\
&\frac{\partial\psi_1}{\partial
t}=\kappa_1\Delta\psi_1+\lambda_1\psi_1-a_1\rho_n\psi_1-\alpha_1|\psi_0|^2\psi_1-a_2|\psi_2|^2\psi_1-
a_3|\psi_1|^2\psi_1,\\
&\frac{\partial\psi_2}{\partial
t}=\kappa_2\Delta\psi_2+\lambda_2\psi_2-b_1\rho_n\psi_2-\alpha_2|\psi_0|^2\psi_2-a_2|\psi_1|^2\psi_2-
b_3|\psi_2|^2\psi_2,\\
&\frac{\partial\rho_n}{\partial
t}=\kappa_3\Delta\rho_n+\lambda_3\rho_n-\frac{\alpha_0}{2}|\psi_0|^2-\frac{a_1}{2}|\psi_1|^2-
\frac{b_1}{2}|\psi_2|^2+c_2\rho^2_n-c_3\rho^3_n,\\
&\frac{\partial }{\partial n} (\psi_0, \psi_1, \psi_2, \rho_n)=0
\qquad \qquad \qquad \text{ on } \partial\Omega,
\end{aligned}
\right.\label{8.252}
\end{equation}
\end{widetext}
where the coefficients satisfy that  for any $0\leq i\leq 3$  and 
$1\leq j\leq 3$, 
$$\alpha_i>0, \quad  a_j>0, \quad 
b_1,b_3,c_2,c_3>0.$$

Equations (\ref{8.252}) should be the same as (\ref{8.230}) for $H=0$.
Therefore we assume that when $H=0$,
\begin{equation}
\kappa_0=\kappa_1,\ \ \ \ \lambda_0=\lambda_1,\ \ \ \
\alpha_0=a_1,\ \ \ \ \alpha_1=0,\ \ \ \ \alpha_2=a_2,\ \ \ \
\alpha_3=a_3.\label{8.253}
\end{equation}

Based on the physical facts, we also assume that
\begin{equation}
\begin{aligned}
& 
\lambda_0=\lambda_1(T,p)+\widetilde{\lambda}(T,p,H),\\
& \widetilde{\lambda}(T,p,H)>0 &&  \text{if}\ H\neq 0,\\
& \widetilde{\lambda}(T,p,H)\rightarrow 0 &&  \text{if}\
H\rightarrow 0.
\end{aligned}
\label{8.254}
\end{equation}

When the magnetic field $H$ and the pressure $p$ are homogeneous on
$\Omega$, the problem (\ref{8.252}) can be reduced to
\begin{equation}
\begin{aligned}
&\frac{d\rho_0}{dt}=\lambda_0\rho_0-\alpha_0\rho_n\rho_0-\alpha_1\rho_1\rho_0-\alpha_2\rho_2\rho_0-
\alpha_3\rho^2_0,\\
&\frac{d\rho_1}{dt}=\lambda_1\rho_1-a_1\rho_n\rho_1-\alpha_1\rho_0\rho_1-a_2\rho_2\rho_1-a_3\rho^2_1,\\
&\frac{d\rho_2}{dt}=\lambda_2\rho_2-b_1\rho_n\rho_2-\alpha_2\rho_0\rho_2-a_2\rho_2\rho_1-b_3\rho^2_2,\\
&\frac{d\rho_n}{dt}=\lambda_3\rho_n-\frac{\alpha_0}{2}\rho_0-\frac{a_1}{2}\rho_1-\frac{b_1}{2}\rho_2 +
c_2\rho^2_n-c_3\rho^3_n,
\end{aligned}
\label{8.255}
\end{equation}
where $\rho_i=|\psi_i|^2$  $(i=0,1,2)$, and $\lambda_i$  $(i=1,2,3)$ are
as in (\ref{8.233}).

Let $\lambda_1,\lambda_2$, and $\lambda_3$ be that as shown in
Figure \ref{f8.39}(a)-(c) respectively. Then, due to
(\ref{8.254}) the curves $\lambda_j(T,p)=0(0\leq j\leq 3)$ in
$PT$-plane are schematically illustrated by Figure \ref{f8.44}.
\begin{figure}[hbt]
  \centering
  \includegraphics[width=0.2\textwidth]{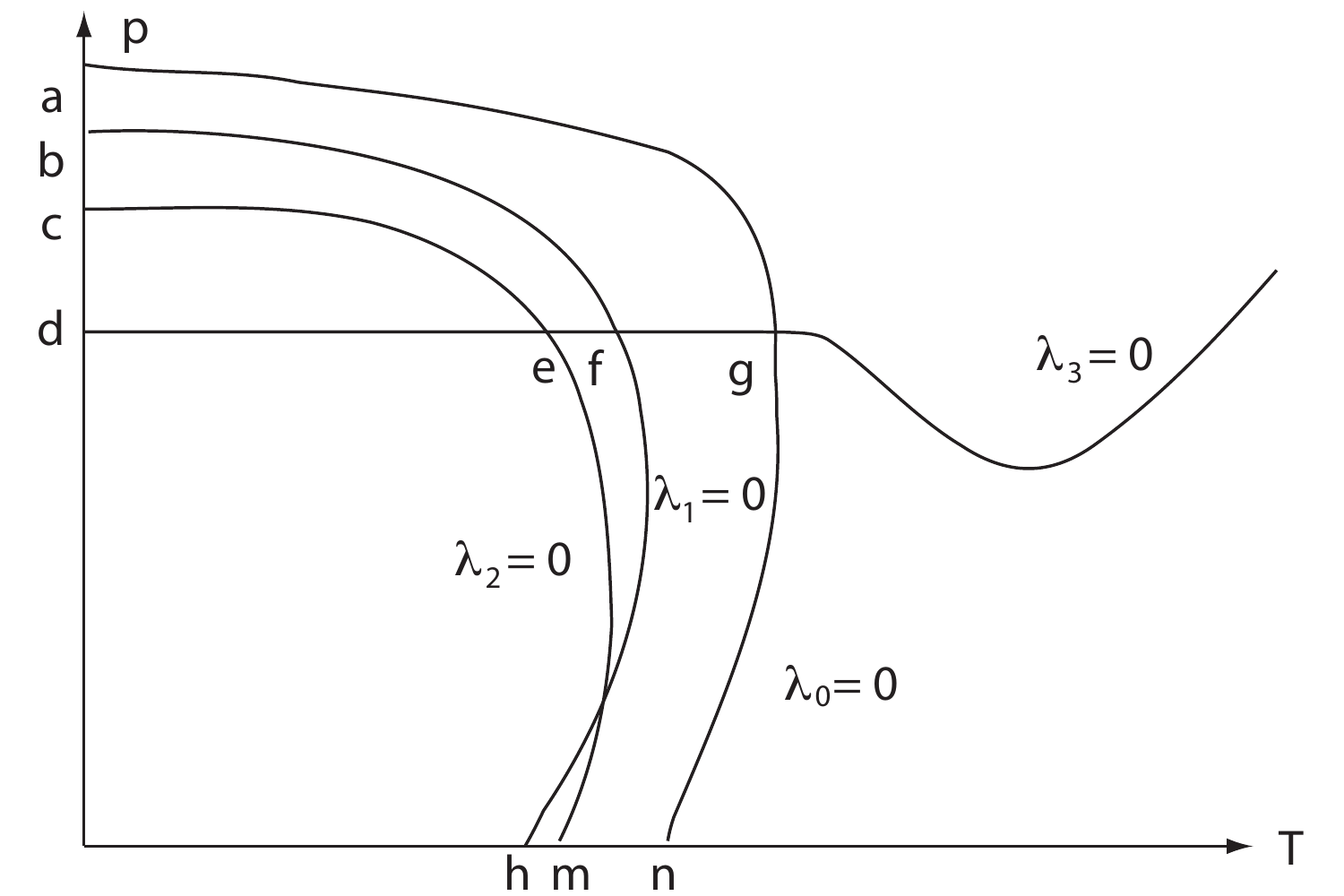}
  \caption{Curve $\widehat{agn}$ is $\lambda_0=0,
\widehat{bfh}$ is $\lambda_1=0, \widehat{cem}$ is $\lambda_2=0$,
and $\widehat{defg}$ is $\lambda_3=0$.}\la{f8.44}
 \end{figure}

Let the applied magnetic field $H\neq 0$ such that
\begin{equation}
0<\widetilde{\lambda}(T,p,H)<\varepsilon ,\ \ \ \ \text{for}\
\varepsilon >0\ \text{small},\label{8.256}
\end{equation}
where $\widetilde{\lambda}$ is as in (\ref{8.254}). Then, under
the conditions
(\ref{8.237}),(\ref{8.238}),(\ref{8.253}),(\ref{8.254}) and
(\ref{8.256}), by using the same fashion as in Theorem \ref{t8.15} and
\ref{t8.16}, we can prove the following transition theorem for
(\ref{8.255}).

\bt\la{t8.18}
Assume the conditions
(\ref{8.237}),(\ref{8.238}),(\ref{8.253}),(\ref{8.254}) and
(\ref{8.256}), then for $H\neq 0$ there exist two curve segments
$\widehat{f^{\prime}h^{\prime}}$ near $\lambda_1=0$ and
$\widehat{e^{\prime}m^{\prime}}$ near $\lambda_2=0$ in  the $PT$-plane
as shown in Figure \ref{f8.45} such that the following assertions hold true:

\begin{itemize}
\item[(1)] The system (\ref{8.255}) has a transition in curve
segment $\lambda_0=0$ with $\lambda_3<0$ (i.e., the curve segment
$\widehat{gn}$ in Figure \ref{f8.45}), which is Type-I for
$\alpha^2_0-2|\lambda_3|\alpha_3\leq 0$, and is Type-II for
$\alpha^2_0-2|\lambda_3|\alpha_3>0$. The transition solution is
given by $(\rho^*_0,0,0,\rho^*_n)$ with $\rho^*_0>0, \rho^*_n<0.$

\item[(2)] The system has a second transition from
$(\rho^*_0,0,0,\rho^*_n)$ in the curve segment
$\widehat{f^{\prime}h^{\prime}}$ (i.e.,
$\lambda_1-a_1\rho^*_n-\alpha_1\rho^*_0=0)$, and the transition
solution is as
$(\rho^{\prime}_0,\rho^{\prime}_1,0,\rho^{\prime}_n)$ with
$\rho^{\prime}_0>0,\rho^{\prime}_1>0$ and $\rho^{\prime}_n<0$.
\item[(3)]\ The system has a third transition from
$(\rho^{\prime}_0,\rho^{\prime}_1,0,\rho^{\prime}_n)$ in the curve
segment $\widehat{e^{\prime}m^{\prime}}$ (i.e.,
$\lambda_2-b_1\rho^{\prime}_n-\alpha_2\rho^{\prime}_0-a_2\rho^{\prime}_1=0)$,
and the transition solution is
$(\rho^{\prime\prime}_0,\rho^{\prime\prime}_1,\rho^{\prime\prime}_2,\rho^{\prime\prime}_n)$
with $\rho^{\prime\prime}_i>0 (0\leq i\leq 2)$ and
$\rho^{\prime}_n<0$.
\end{itemize}
\et

\br\la{r8.13} 
{\rm
The first transition of (\ref{8.255}) in
curve segment $\widehat{gn}$ corresponds to the phase transition
of $^3$He in a magnetic field between the normal liquid and
superfluid phase $A_1$, and the second transition in
$\widehat{f^{\prime}h^{\prime}}$ corresponds to the phase
transition between superfluid phase $A_1$ and $A$, and the third
transition in $\widehat{e^{\prime}m^{\prime}}$ corresponds to the
phase transition between superfluid phases $A$ and $B$; see Figure
\ref{f8.45}.
}\er

\br\la{r8.14}
{\rm 
The transition theorems, Theorems \ref{t8.14}-\ref{t8.16}
and \ref{t8.18}, provide theoretical foundation to explain the $PT$-phase
diagrams of superfluidity, meanwhile they support these models of
liquid He which are based on the phenomenology.
}
\er

\begin{figure}[hb]
  \centering
  \includegraphics[width=0.2\textwidth]{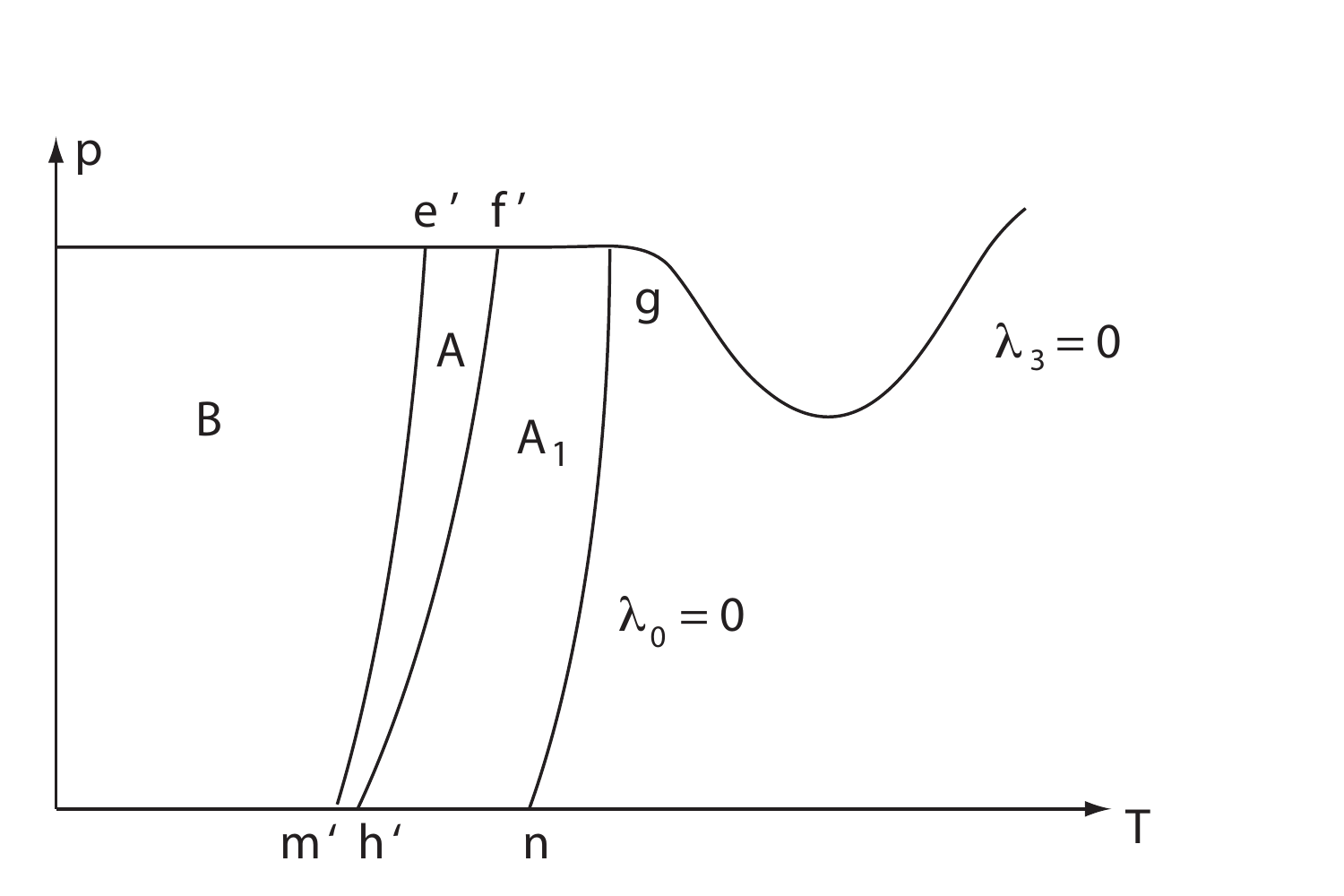}
  \caption{}\la{f8.45}
 \end{figure}

\section{Physical Remarks}
By carefully examining the classical phase transition diagrams and with both mathematical and physical insights offered by the dynamical transition theory, we 
derived two new models for superfluidity of $^3$He with or without applied field. A crucial component of these two models is the introduction of three wave functions  to represent Anderson-Brinkman-Morel (ABM)  and the Balian-Werthamer (BM)  states.

Then we have obtained   a theoretical  $PT$-phase diagram of $^3$He as shown in Figure~\ref{f8.41}, based on the models and the dynamic phase transition analysis. A few main characteristics of the results are as follows.

First, the analysis shows the existence of a unstable region  $H=H_1 \cup H_2$, in which  the solid state and the superfluid states $A$ and $B$ appear randomly depending on fluctuations.   In particular, in $H_1$, phase $B$ superfluid state and the solid may appear, and in $H_2$ the phase $A$ superfluid state and the solid sate may appear.

Second, theoretical analysis suggests the existence of phase C superfluid state, which is characterized by the wave function $\psi_2$, representing $|\uparrow\downarrow >+|\downarrow\uparrow>$. However, phase C region is very narrow, which may be the reason why it is hard to be observed in experiments.

Third, the curve $\widehat{bcd}$ is the first critical curve where phase transition between normal fluid and superfluid states occur.
The curve $\widehat{f'c}$
   is the coexistence curve between phases $A$ and $B$. The curve $\widehat{bc}$ is the coexistence curve between normal fluid state and the phase $A$ superfluid state, the curve $\widehat{cd}$ is the coexistence curve between normal fluid state and the phase $C$ superfluid state, 
  the curve $\widehat{ch'}$ is the coexistence curve between  the phases $B$ and $C$ superfluid states.
  
  Fourth, Theorems~\ref{t8.15}-\ref{t8.17}  imply that near the two triple points $b$  and $c$, there is a possibility of the existence of two switch points, where the transition on the corresponding coexistence curve switches types at each switch point. The existence of such switch points depends on the physical parameters.

  In comparison to classical results as shown in Figure~\ref{f8.38}, our results lead to the predictions  of the existence of 1)   an unstable region $H$, 2)  a new phase C
in a narrow region, and 3) switch points.   It is hoped that these predictions 
  will be useful for designing better physical experiments and lead to better understanding of the physical mechanism of superfluidity.

\appendix
  \section{General Principles of Phase Transition Dynamics}
In this appendix,  we introduce a new phase dynamic transition classification scheme 
to classify phase transitions into three categories: Type-I, Type-II and Type-III, corresponding mathematically  continuous,  jump, and  mixed transitions, respectively. 

\subsection{Dynamic transition theory}
In sciences, nonlinear dissipative systems are generally governed
by differential equations, which can be expressed in the following
abstract form 
Let $X$  and $ X_1$ be two Banach spaces,   and $X_1\subset X$ a compact and
dense inclusion. Hereafter we always consider the following
nonlinear evolution equations
\begin{equation}
\frac{du}{dt}=L_{\lambda}u+G(u,\lambda),\qquad u(0)=\varphi ,
\label{5.1}
\end{equation}
where $u:[0,\infty )\rightarrow X$ is unknown function,  and 
$\lambda\in \R^1$  is the system parameter.

Assume that $L_{\lambda}:X_1\rightarrow X$ is a parameterized
linear completely continuous field depending contiguously on
$\lambda\in \R^1$, which satisfies
\begin{equation}
\left. 
\begin{aligned} 
&L_{\lambda}=-A+B_{\lambda}   && \text{a sectorial operator},\\
&A:X_1\rightarrow X   && \text{a linear homeomorphism},\\
&B_{\lambda}:X_1\rightarrow X&&  \text{a linear compact  operator}.
\end{aligned}
\right.\label{5.2}
\end{equation}
In this case, we can define the fractional order spaces
$X_{\sigma}$ for $\sigma\in \R^1$. Then we also assume that
$G(\cdot ,\lambda ):X_{\alpha}\rightarrow X$ is $C^r(r\geq 1)$
bounded mapping for some $0\leq\alpha <1$, depending continuously
on $\lambda\in \R^1$, and
\begin{equation}
G(u,\lambda )=o(\|u\|_{X_{\alpha}}) \qquad  \forall\lambda\in
\R^1.\label{5.3}
\end{equation}

Hereafter we always assume the conditions (\ref{5.2}) and
(\ref{5.3}), which represent that the system (\ref{5.1}) has
a dissipative structure.

A state of the system (\ref{5.1}) at $\lambda$ is usually referred to as a compact invariant set $\Sigma_{\lambda}$. In many applications, 
$\Sigma_{\lambda}$ is a singular point or a periodic orbit. A state $\Sigma_{\lambda}$
of (\ref{5.1}) is stable if $\Sigma_{\lambda}$ is an attractor;
otherwise $\Sigma_{\lambda}$ is called unstable.

\begin{defi}
\label{d7.1}
We say that the system (\ref{5.1}) has a
phase transition from a state $\Sigma_{\lambda}$ at $\lambda
=\lambda_0$ if $\Sigma_{\lambda}$ is stable on $\lambda <\lambda_0$
(or on $\lambda >\lambda_0$) and is unstable on $\lambda
>\lambda_0$ (or on $\lambda <\lambda_0$).  The critical parameter $\lambda_0$ is
called a critical point.
 In other words, the
phase transition corresponds to an exchange of stable states.
\end{defi}

Obviously, the attractor bifurcation of (\ref{5.1}) is a type of
transition. However,  bifurcation and
transition are two different, but related concepts. 

Let $\{\beta_j(\lambda )\in \C\ \   |\ \ j \in \N\}$  be the eigenvalues (counting multiplicity) of $L_{\lambda}$, and  assume that
\begin{align}
&  \text{Re}\ \beta_i(\lambda )
\left\{ 
 \begin{aligned} 
 &  <0 &&    \text{ if } \lambda  <\lambda_0,\\
& =0 &&      \text{ if } \lambda =\lambda_0,\\
& >0&&     \text{ if } \lambda >\lambda_0,
\end{aligned}
\right.   &&  \forall 1\leq i\leq m,  \label{5.4}\\
&\text{Re}\ \beta_j(\lambda_0)<0 &&  \forall j\geq
m+1.\label{5.5}
\end{align}

The following theorem is a basic principle of transitions from
equilibrium states, which provides sufficient conditions and a basic
classification for transitions of nonlinear dissipative systems.
This theorem is a direct consequence of the center manifold
theorems and the stable manifold theorems; we omit the proof.

\bt\la{t5.1}
 Let the conditions (\ref{5.4}) and
(\ref{5.5}) hold true. Then, the system (\ref{5.1}) must have a
transition from $(u,\lambda )=(0,\lambda_0)$, and there is a
neighborhood $U\subset X$ of $u=0$ such that the transition is one
of the following three types:

\begin{itemize}
\item[(1)] {\sc Continuous Transition}: 
there exists an open and dense set
$\widetilde{U}_{\lambda}\subset U$ such that for any
$\varphi\in\widetilde{U}_{\lambda}$,  the solution
$u_{\lambda}(t,\varphi )$ of (\ref{5.1}) satisfies
$$\lim\limits_{\lambda\rightarrow\lambda_0}\limsup_{t\rightarrow\infty}\|u_{\lambda}(t,\varphi
)\|_X=0.$$ 

\item[(2)] {\sc Jump Transition}: 
for any $\lambda_0<\lambda <\lambda_0+\varepsilon$ with some $\varepsilon >0$, there is an open
and dense set $U_{\lambda}\subset U$ such that 
for any $\varphi\in U_{\lambda}$, 
$$\limsup_{t\rightarrow\infty}\|u_{\lambda}(t,\varphi
)\|_X\geq\delta >0,$$ 
for some $\delta >0$  independent of $\lambda$.

\item[(3)] {\sc Mixed Transition}: 
for any $\lambda_0<\lambda <\lambda_0+\varepsilon$  with some $\varepsilon >0$, 
$U$ can be decomposed into two open sets
$U^{\lambda}_1$ and $U^{\lambda}_2$  ($U^{\lambda}_i$ not necessarily
connected):
$\bar{U}=\bar{U}^{\lambda}_1+\bar{U}^{\lambda}_2$, $U^{\lambda}_1\cap U^{\lambda}_2=\emptyset$,  such that
\begin{align*}
&\lim\limits_{\lambda\rightarrow\lambda_0}\limsup_{t\rightarrow\infty}\|u(t,\varphi
)\|_X=0   &&   \forall\varphi\in U^{\lambda}_1,\\
& \limsup_{t\rightarrow\infty}\|u(t,\varphi
)\|_X\geq\delta >0 && \forall\varphi\in U^{\lambda}_2.
\end{align*}
\end{itemize}
\et

With this theorem in our disposal, we are in position to give a new dynamic classification scheme for dynamic phase transitions.

\begin{defi}
The phase transitions for  (\ref{5.1}) at $\lambda =\lambda_0$ is classified using  their  dynamic properties: continuous, jump, and mixed as given in Theorem~\ref{t5.1}, which are called Type-I, Type-II and Type-III respectively.
\end{defi}

An important aspect of the  transition theory is to determine which 
of the three types of transitions given by Theorem \ref{t5.1} occurs in
a specific  problem. A corresponding dynamic transition  theory has been developed recently by the authors for this purpose; see \cite{chinese-book}. We refer interested readers to these references for details of the theory. 

\subsection{New Ginzburg-Landau models for equilibrium phase transitions}
\label{s7.2.2}
In this section, we recall a new time-dependent Ginzburg-Landau theory for modeling  equilibrium phase transitions in statistical physics. 

Consider a thermal system with a control  parameter $\lambda$. 
By the mathematical characterization of gradient systems and the le Ch\^atelier principle, for a system with
thermodynamic potential ${\mathcal{H}}(u,\lambda )$, the governing
equations are essentially determined by the functional
${\mathcal{H}}(u,\lambda )$.
When the order parameters $(u_1,\cdots,u_m)$ are nonconserved
variables, i.e., the integers
$$\int_{\Omega}u_i(x,t)dx=a_i(t)\neq\text{constant}.$$
then the time-dependent equations are given by
\begin{equation}
\left.
\begin{aligned} 
&\frac{\partial u_i}{\partial
t}=-\beta_i\frac{\delta}{\delta u_i}{\mathcal{H}}(u,\lambda
)+\Phi_i(u,\nabla u,\lambda ),
\end{aligned}
\right.\label{7.30}
\end{equation}
for  $1 \le i \le m$, where $\beta_i>0$ and $\Phi_i$ satisfy
\begin{equation}
\int_{\Omega}\sum_i\Phi_i\frac{\delta}{\delta
u_i}{\mathcal{H}}(u,\lambda )dx=0.\label{7.31}
\end{equation}
The condition (\ref{7.31})  is  required by
the Le Ch\^atelier principle. In the concrete problem, the terms
$\Phi_i$ can be determined by physical laws and (\ref{7.31}). We remark here that following the le Ch\^atelier principle, one should have an inequality constraint. However   physical systems often obey most simplified rules, as  many existing models for specific problems are consistent with the equality constraint here. This remark applies to the constraint (\ref{7.37}) below as well.

When the order parameters are the number density and the system
has no material exchange with the external, then $u_j$  $(1\leq j\leq
m)$ are conserved, i.e.,
\begin{equation}
\int_{\Omega}u_j(x,t)dx=\text{constant}.\label{7.32}
\end{equation}
This conservation law requires a continuity equation
\begin{equation}
\frac{\partial u_j}{\partial t}=-\nabla\cdot J_j(u,\lambda
),\label{7.33}
\end{equation}
where $J_j(u,\lambda )$ is the flux of component $u_j$, satisfying
\begin{equation}
J_j=-k_j\nabla (\mu_j-\sum_{i\neq j}\mu_i),\label{7.34}
\end{equation}
where $\mu_l$ is the chemical potential of component $u_l$, 
\begin{equation}
\mu_j-\sum_{i\neq j}\mu_i=\frac{\delta}{\delta
u_j}{\mathcal{H}}(u,\lambda )-\phi_j(u,\nabla u,\lambda
), \label{7.35}
\end{equation}
and  $\phi_j(u,\lambda )$ is a function depending on the other
components $u_i$ $(i\neq j)$. Thus, from
(\ref{7.33})-(\ref{7.35}) we obtain the dynamical equations as
follows
\begin{equation}
\begin{aligned} 
&\frac{\partial u_j}{\partial
t}=\beta_j\Delta\left[\frac{\delta}{\delta
u_j}{\mathcal{H}}(u,\lambda )-\phi_j(u,\nabla u,\lambda )\right],
\end{aligned}
\label{7.36}
\end{equation}
for $1 \le j \le m$, where $\beta_j>0$ are constants,  and  $\phi_j$ satisfy
\begin{equation}
\int_{\Omega}\sum_j\Delta\phi_j\cdot\frac{\delta}{\delta
u_j}{\mathcal{H}}(u,\lambda )dx=0.\label{7.37}
\end{equation}

When $m=1$, i.e., the system is a binary system, consisting of
two components $A$ and $B$, then the term $\phi_j=0$. The above model covers the classical Cahn-Hilliard model. It is worth mentioning that for multi-component systems, these $\phi_j$ play an important rule in deriving good time-dependent models.

If the order parameters $(u_1,\cdots,u_k)$ are coupled to the
conserved variables $(u_{k+1},\cdots,u_m)$, then the dynamical
equations are
\begin{equation}
\begin{aligned} 
&\frac{\partial u_i}{\partial t}
   =-\beta_i\frac{\delta}{\delta u_i}{\mathcal{H}}(u,\lambda)+\Phi_i(u,\nabla u,\lambda ),\\
& \frac{\partial u_j}{\partial t}
  =\beta_j\Delta\left[\frac{\delta}{\delta u_j}{\mathcal{H}}(u,\lambda )
    -\phi_j(u,\nabla u,\lambda )\right],\\
\end{aligned}
\label{7.38}
\end{equation}
for  $ 1 \le i \le k$  and $k+1 \le j \le m$.
Here $\Phi_i$  and $\phi_j$ satisfy (\ref{7.31})   and (\ref{7.37}), respectively.

The model (\ref{7.38}) we derive here  gives a general form of the governing
equations to thermodynamic phase transitions, and will play crucial role in studying the dynamics of equilibrium phase transitions in statistic physics.

\bibliographystyle{siam}
\def\cprime{$'$}

\end{document}